\documentclass[pra,twocolumn,notitlepage]{revtex4-2}
\pdfoutput=1
\usepackage[T1]{fontenc}
\usepackage{graphics}
\usepackage{url}
\usepackage[colorlinks=true, urlcolor=blue, citecolor=blue, anchorcolor=red]{hyperref}
\usepackage{color}
\usepackage{physics}
\usepackage{bbm}
\usepackage{soul}
\usepackage{mathtools}
\usepackage{amssymb}
\usepackage{dsfont}
\usepackage{amsmath}
\newcommand{\eq}[2]{\begin{equation}\begin{split} \label{eq:#1} #2 \end{split}\end{equation}}
\DeclareMathOperator*{\argmax}{arg\,max}

\DeclareMathOperator{\arctantwo}{arctan2}
\begin{document}

\title{Hardware-efficient variational quantum algorithms for time evolution}

\author{Marcello Benedetti}
\email{marcello.benedetti@cambridgequantum.com}
\affiliation{Cambridge Quantum Computing Limited, SW1E 6DR London, United Kingdom}

\author{Mattia Fiorentini}
\affiliation{Cambridge Quantum Computing Limited, SW1E 6DR London, United Kingdom}

\author{Michael Lubasch}
\affiliation{Cambridge Quantum Computing Limited, SW1E 6DR London, United Kingdom}

\date{July 22, 2021}

\begin{abstract}
Parameterized quantum circuits are a promising technology for achieving a quantum advantage. An important application is the variational simulation of time evolution of quantum systems. To make the most of quantum hardware, variational algorithms need to be as hardware-efficient as possible. Here we present alternatives to the time-dependent variational principle that are hardware-efficient and do not require matrix inversion. In relation to imaginary time evolution, our approach significantly reduces the hardware requirements. With regards to real time evolution, where high precision can be important, we present algorithms of systematically increasing accuracy and hardware requirements. We numerically analyze the performance of our algorithms using quantum Hamiltonians with local interactions.
\end{abstract}

\maketitle

\section{Introduction}

Small quantum computers are available today and offer the exciting opportunity to explore classically difficult problems for which a quantum advantage may be achievable.
The simulation of time evolution of quantum systems is an example of such problems where the advantage of using a quantum computer over a classical one is well understood~\cite{Wi96, Ll96, AbLl97, Za98, KaEtAl08}.
This simulation is also important for our understanding of quantum chemistry and materials science which are key application areas for future quantum computers~\cite{KaEtAl11, CaEtAl19, McEtal20, BaEtAl20}.
One of the main challenges in the design of time evolution algorithms is to reduce their experimental requirements without sacrificing their accuracy.

Although significant progress has been made based on the original quantum algorithm for simulating time evolution~\cite{Ll96}, this algorithm faces obstacles on current quantum hardware which lacks quantum error correction~\cite{SmEtAl19, GuEtAl19, FaZh20}.
Promising alternatives are variational hybrid quantum-classical algorithms~\cite{PeEtAl14, McEtAl16} and variational quantum simulation~\cite{LiBe17, YuEtAl19}.
In these approaches, a parameterized quantum circuit (PQC) is prepared on a quantum computer and variationally optimized to solve the problem of interest. Promising examples of quantum advantage obtained with PQCs have already been identified for time evolution~\cite{LiEtAl20,cirstoiu2020variational} and in other contexts such as for nonlinear partial differential equations~\cite{LuEtAl18, LuEtAl20}, dynamical mean field theory~\cite{JaEtAl20,rungger2020dynamical}, and machine learning~\cite{BeEtAl19}.

\begin{figure}[!ht]
\centering
\includegraphics[width=70mm]{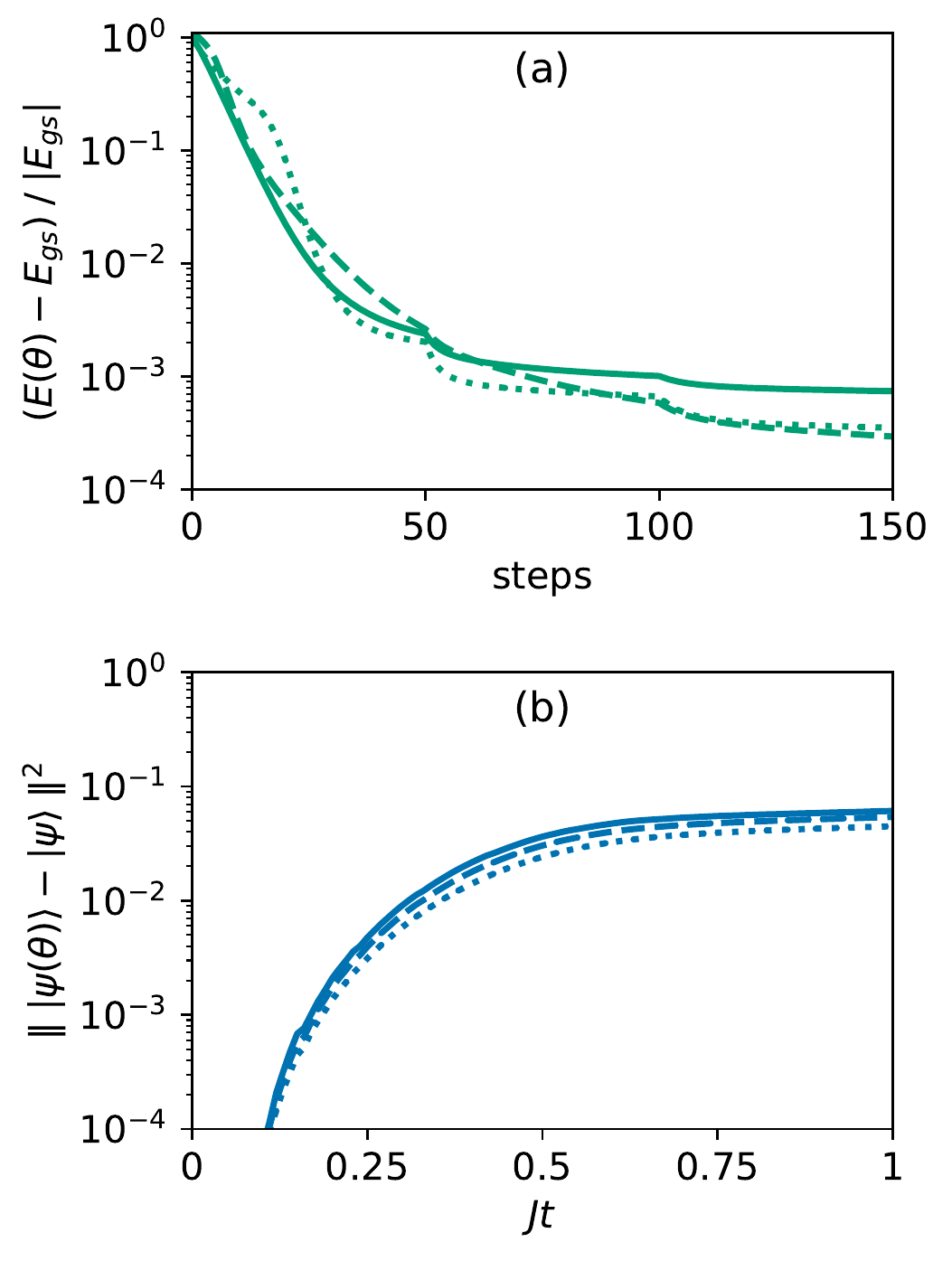}
\vspace{-4mm}
\caption{
Numerical simulation of our hardware-efficient time evolution algorithms. We consider the quantum Ising chain $ \hat{H}= -J \left( \sum_{j=1}^{n-1} \hat{\sigma}^z_j \hat{\sigma}^z_{j+1} + \lambda \sum_{j=1}^n \hat{\sigma}^x_j \right)$ with $J=1$, $\lambda=0.2$ and length $n = 8$ (dotted line), $10$ (dashed line), and $12$ (solid line). We use a PQC Ansatz of depth $\mathcal{D} = 2$, which is the smallest nontrivial Ansatz among those considered in this work. The resulting causal cones require quantum hardware with only six qubits. 
(a) Imaginary time evolution of a randomly initialized PQC. We use \textsc{angle update} with one sweep for $50$ time steps of size $0.05$, then $50$ time steps of size $0.03$, and then $50$ time steps of size $0.01$. $E(\theta)$ is the variational state's energy and $E_{gs}$ is the exact ground state energy. The algorithm achieves relative energy errors below $10^{-3}$.
(b) Real time evolution of the initial state $\ket{00\ldots0}$. We use \textsc{cone update} with six sweeps and time step size $0.01$. We plot the squared Euclidean distance between the variational state $\ket{\psi(\boldsymbol{\theta})}$ and the exact evolved state $\ket{\psi}$ versus time $t$ (in units of $1/J$). A $0.05$ distance is obtained at the end of the simulation.
}
\label{fig:results}
\end{figure}

In this paper, we focus on making the variational simulation of time evolution as efficient as possible in terms of quantum hardware resources.
Existing proposals are based on the time-dependent variational principle or variants thereof~\cite{LiBe17, YuEtAl19}. These approaches require matrix inversion which poses computational challenges for ill-conditioned matrices.
Our contribution is a set of alternative techniques that do not need matrix inversion and that allow one to systematically increase the simulation accuracy, together with the hardware requirements, according to experimental capabilities.
We analyze the hardware/accuracy trade-off and show that for imaginary time evolution our algorithm significantly reduces the hardware requirements over existing methods and produces accurate ground state approximations.
For real time evolution, the accuracy per time step is essential. We present a hierarchy of algorithms where the accuracy can be systematically improved by utilizing more hardware resources.
Figure~\ref{fig:results} illustrates the performance of some of the algorithms developed in this paper.

Our strategy is inspired by several tensor network concepts.
Firstly, we apply the Trotter product formula to the time evolution operator and optimize the Ansatz one Trotter term after another.
A similar procedure is used in the time-evolving block decimation algorithm~\cite{Vi03, Vi04, DaEtAl04, Vi07} as well as with projected entangled pair states~\cite{JoEtAl08, LuCiBa14a, LuCiBa14b}.
Secondly, for a given Trotter term, we restrict the optimization to its causal cone. This is an important concept for the multiscale entanglement renormalization Ansatz~\cite{Vi08, EvGi09} as well as for matrix product states~\cite{Ha09}. The same concept is used in the design of noise-robust quantum circuits~\cite{KiSw17} and to simulate infinite matrix product states on a quantum computer~\cite{BarEtAl20}. Thirdly, we perform the optimization coordinatewise~\cite{ViTh18, NaFuTo19, PaIoOzMc19, ostaszewski2019quantum}, i.e., we optimize a set of parameters at a time while keeping all the others fixed. A similar approach is widely used in tensor network optimization where tensors are optimized one after another~\cite{VeMuCi08, Or14}.

In this work, we call \emph{hardware-efficient} any variational quantum algorithm that (i) can use parameterized quantum circuits tailored to the physical device~\cite{Kandala_2017} or (ii) exploits other concepts that reduce the number of qubits or gates~\cite{OlEtAl20, TaEtAl21} such as causal cones. There exist alternative efficient schemes based on measuring and resetting qubits during computation~\cite{huggins2019towards, FoEtAl20, ChEtAl21} and for periodic quantum systems~\cite{liu2020simulating,manrique2020momentumspace}.

The paper has the following structure. Section~\ref{sec:Methods} summarizes our methods, Sec.~\ref{sec:Results} presents our mathematical and numerical results, and Sec.~\ref{sec:Discussion} contains our concluding discussion. The technical details are in the appendices. In a companion paper~\cite{amaro2021filtering} we apply our algorithm to combinatorial optimization problems of up to 23 qubits.

\section{Methods}
\label{sec:Methods}

\subsection{Taking apart time evolution}

In the following, we focus on real time evolution. To obtain imaginary time evolution one simply needs to substitute $t$ by $-it$.

The simulation of real time evolution consists of approximating the action of the operator $e^{-i t \hat{H}}$ on an initial state $\ket{\psi_0}$ of $n$ qubits. The Hamiltonian is assumed to have the general form $\hat{H} = \sum_{k=1}^K h_k \hat{H}_k$, where $\hat{H}_k$ are tensor products of Pauli operators, $h_k$ are real numbers, and $K \sim O(\text{poly}(n))$. We approximate the evolution by a sequence of short-time evolutions using the well-known Trotter product formula. With $N$ time steps of size ${ \tau=t/N }$ one obtains $e^{-i t \hat{H}} \approx U(\tau)^N$, where ${ U(\tau) = e^{-i \tau h_K \hat{H}_K} \cdots e^{-i \tau h_1 \hat{H}_1} }$. The accuracy of this approximation can be improved using higher-order Trotter product formulas~\cite{HaSu05}.

We variationally simulate the sequence of Trotter terms one term at a time.
Consider the $k$th term, $e^{-i \tau h_k \hat{H}_k}$, and let $\ket{\psi_{k-1}}$ be the variational approximation to the previous step. We minimize the squared Euclidean distance $\Vert \ket{\psi(\boldsymbol{\theta})} - e^{-i \tau h_{k} \hat{H}_{k}} \ket{ \psi_{k-1} }\Vert^{2}$ via the variational parameters $\boldsymbol{\theta}$ by maximizing the objective function:
\eq{main_objective}{
\mathcal{F}_k(\boldsymbol{\theta}) &= \Re \left( \bra{\psi_{k-1}} e^{i \tau h_k \hat{H}_k} \ket{\psi(\boldsymbol{\theta})} \right) ,
}
where $\text{Re}(\cdot)$ denotes the real part of a complex number, see Appendix~\ref{a:variational} for details. This optimization is carried out for all $K$ terms in $U(\tau)$. The process is repeated $N$ times, after which the simulation of time evolution is completed. 

\begin{figure*}
\centering
\includegraphics[width=164.617mm]{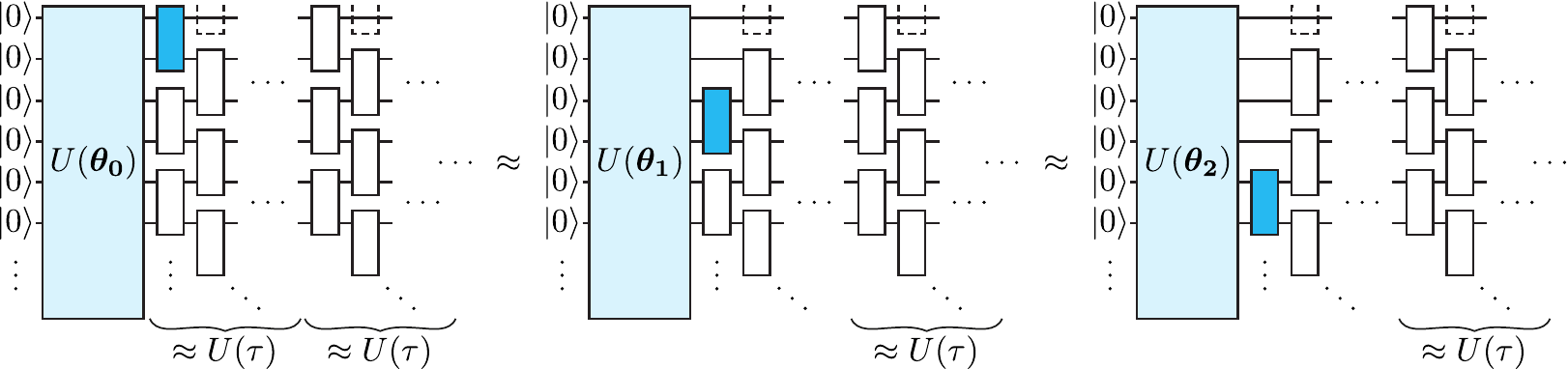}
\vspace{-2mm}
\caption{
The Trotter product formula applied to the time evolution operator results in repeated products of $U(\tau)$, where $\tau$ is the time step, acting on the initial state $\ket{\psi_0} = U(\boldsymbol{\theta_{0}}) \ket{\boldsymbol{0}}$. We approximate time evolution one term after another, each time finding the optimal variational parameters $\boldsymbol{\theta_{1}}, \boldsymbol{\theta_{2}},\ldots$}
\label{fig:Trotter}
\end{figure*}

Figure~\ref{fig:Trotter} illustrates a few steps of the method. The variational state consists of a PQC (light blue rectangle) acting on the $\ket{\boldsymbol{0}} \equiv \ket{0}^{\otimes n}$ state of the computational basis. At each step, a term of the Trotter formula is selected (blue rectangle) and simulated by the variational method.

\begin{figure*}
\centering
\includegraphics[width=148.229mm]{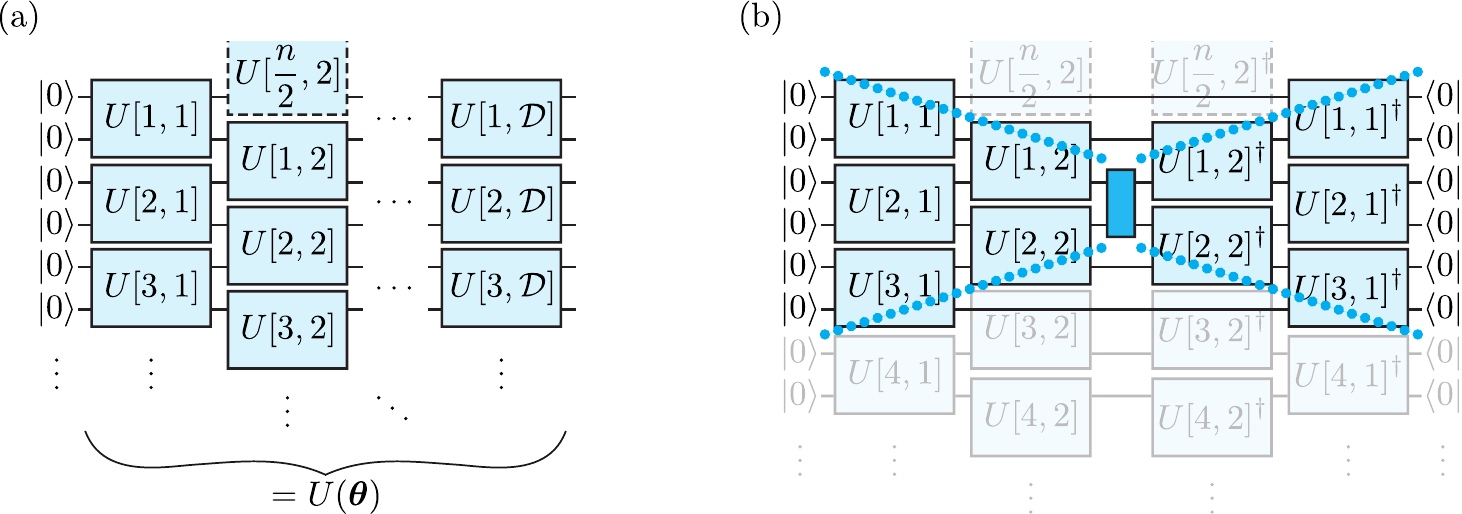}
\vspace{-2mm}
\caption{\label{fig:PQCAndCones}
(a) PQC Ansatz for $n$ qubits and depth $\mathcal{D}$. We consider open or periodic boundary conditions. For periodic boundary conditions the unitary block $U[\frac{n}{2},2]$ connects the first and the last qubit. (b) The computation of an expectation simplifies if one considers only the blocks inside the causal cone. We optimize the objective function in Eq.~\eqref{eq:main_objective} using only the variational parameters inside the causal cone. This technique enables us to attack problems of size larger than that of the available quantum hardware.}
\end{figure*}

\subsection{Parameterized quantum circuits and causal cones}
\label{s:pqc_cones}

We consider PQC Ans\"{a}tze composed of generic 2-qubit unitaries acting on nearest neighbors, as shown in Fig.~\ref{fig:PQCAndCones}~(a). The required $1$d qubit-to-qubit connectivity matches that of many existing quantum computers. These Ans\"{a}tze are universal for quantum computation if we adjust their depth accordingly. We refer to the unitaries $U[i,j]$ (light blue rectangles) as \textit{blocks}. In practice, they are made of gates from the hardware's gate set.

An interesting property is that the expectation of an operator acting on a few qubits depends only on the blocks inside its causal cone. Figure~\ref{fig:PQCAndCones}~(b) illustrates an example of causal cone for a two-qubit operator. The expectation can be estimated with a quantum circuit of six qubits and five blocks, regardless of the overall size of the PQC Ans\"{a}tze.

We perform the optimization of the objective function in Eq.~\eqref{eq:main_objective} using only the blocks inside the causal cone of the Trotter term. Thus our variational algorithm requires just the preparation of circuits on quantum hardware of a size restricted by the causal cone. This allows us to work with variational states of size greater than the size allowed by the quantum computer. For example, periodic boundary conditions can be included with no additional hardware requirements. In Fig.~\ref{fig:PQCAndCones}~(a), block $U[\tfrac{n}{2},2]$ operates on the first and last qubits. It is easy to see that such a physical qubit-to-qubit connectivity is not required when using causal cones. The number of required physical qubits depends only on the depth of the Ansatz and on the Trotter term. For long-range operators or operators acting on more than two qubits one can obtain several causal cones.

Throughout this paper we consider the Ansatz in Fig.~\ref{fig:PQCAndCones}~(a) with the first nontrivial depth $\mathcal{D}=2$.
The performance of our algorithms could be systematically improved by successively increasing the depth $\mathcal{D}$.
For example, when the variational error becomes too large during the time evolution, we can include a new column of blocks initialized to $\mathds{1}$ at the end of the PQC to increase its expressive power.
In this way, our Ansatz can be dynamically adapted to guarantee a certain variational error during the entire evolution, analogous to how this is commonly done in the time-evolving block decimation algorithm~\cite{DaEtAl04}.

\begin{figure*}
\centering
\includegraphics[width=172mm]{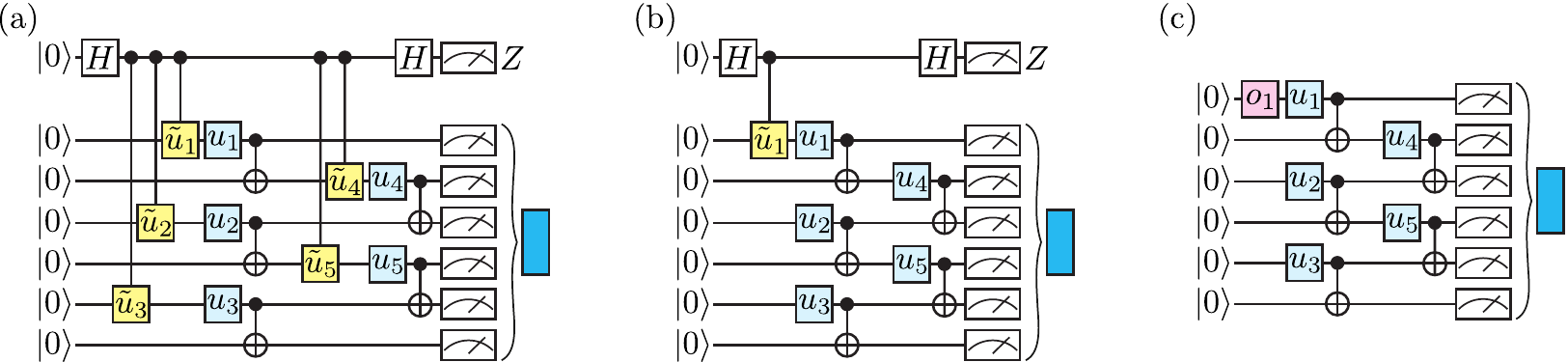}
\caption{
Hadamard tests required for our algorithms. For illustrative purposes, we show a simple PQC Ansatz made of five CNOTs and five parameterized gates $u_i$ (light blue). (a) In \textsc{cone update}, the previous and current PQCs differ by at most five gates. The test requires only local transformations such as the controlled-$\tilde{u}_i$ (yellow). (b) In \textsc{angle update}, the two PQCs differ by one gate. (c) \textsc{angle update} can be further simplified by removing the ancilla qubit and the controlled operation, and introducing an operation $o_i$ (pink). This further simplification is presented in Appendix~\ref{a:hardware-efficient}.}
\label{fig:FHUAndHEHU}
\end{figure*}

\subsection{The coordinatewise update rule}
\label{s:rotosolve}

Let us specialize our discussion to PQCs of the form $U(\boldsymbol{\theta}) = U_D \cdots U_1$ where each gate is either fixed, e.g., a CNOT, or parameterized as ${ U_d = \exp(-i \theta_d G_d) }$, where $\theta_d \in (-\pi, \pi]$ and $G_d$ is a Hermitian and unitary matrix such that $G_d^2=I$. This standard parametrization has nice properties that we exploit to design optimization algorithms.
In Refs.~\cite{ViTh18, NaFuTo19, PaIoOzMc19, ostaszewski2019quantum}, the authors showed that when all parameters but one are fixed the energy expectation value has sinusoidal form. Therefore, there is an analytic expression for the extrema. Here we show that the same is true for the objective function in Eq.~\eqref{eq:main_objective}.

We define the coordinatewise objective for the $d$th parameter as ${ f_{k,d} (x) \equiv \mathcal{F}_k(\theta_1, \cdots, \theta_{d-1}, x, \theta_{d+1}, \cdots, \theta_D) }$, where all parameters but one are fixed to their current value. In Appendix~\ref{a:sinusoidal}, we show that this is a sinusoidal function $f_{k,d} (x) = A_{k,d} \sin(x + B_{k,d} )$ with amplitude $A_{k,d}$ and phase $B_{k,d}$. Thus one can use a coordinatewise optimization procedure as in Refs.~\cite{ViTh18, NaFuTo19, PaIoOzMc19, ostaszewski2019quantum} that neither requires the gradient nor the Hessian. The procedure sweeps through all parameters and sets each of them to their locally optimal value $x^* = \tfrac{\pi}{2} - B_{k,d}$. At $x^*$ the coordinatewise objective attains \mbox{its maximum value of $A_{k,d}$.} 

The estimation of $A_{k,d}$ and $B_{k,d}$ is efficient and leads to a simple update rule. For the $d$th parameter and for the $k$th term we have:
\eq{update_rule}{
\theta_d \leftarrow \tfrac{\pi}{2} - \arctantwo \Big( f_{k,d}(\theta_{d}) , \; f_{k,d}(\theta_d + \tfrac{\pi}{2}) \Big) + \theta_d ,
}
where $\theta_d$ in the right side is the current value of the parameter. This formula requires evaluating the objective at $x=\theta_d$ and $x=\theta_d + \tfrac{\pi}{2}$. However, in Appendix~\ref{a:rotosolve}, we show that $f_{k,d}(\theta_{d})= A_{k,d-1}$ is known from the previous step. Thus the method finds the maxima with a single evaluation, $f_{k,d}(\theta_d + \tfrac{\pi}{2})$.

\subsection{Hardware-efficient implementation}
\label{s:hardware_efficient}

In general, our method optimizes a new PQC for each Trotter term. Let us denote the $(k-1)$th quantum state as $\ket{\psi_{k-1}} = V \ket{\mathbf{0}}$, and the $k$th state as $\ket{\psi(\boldsymbol{\theta})} = U \ket{\mathbf{0}}$, where $V$ and $U$ are PQCs. The objective in Eq.~\eqref{eq:main_objective} and the update rule in Eq.~\eqref{eq:update_rule} can be estimated using a well-known primitive called the Hadamard test.

The Hadamard test can be challenging to execute on hardware when $U$ and $V$ are unrelated quantum circuits due to the potentially large number of controlled operations. The process can be largely simplified if $U$ and $V$ differ only locally, e.g., in few gates or circuit regions. For instance, if one circuit can be efficiently transformed into the other using local adjoints of gates, then the Hadamard test consists of a rather simple quantum circuit. Figure~\ref{fig:FHUAndHEHU}~(a) shows an example with five variational parameters. Gates $u_i$ represent $U$. Gates controlled-$\tilde{u}_i$ represent the local transformations taking $U$ to $V$. The subspaces where the ancilla (top qubit) is $\ket{0}$ and $\ket{1}$ contain $U \ket{\mathbf{0}}$ and $V \ket{\mathbf{0}}$, respectively. Finally, a measurement yields the quantity in Eq.~\eqref{eq:main_objective}. This is discussed in detail in Appendix~\ref{a:hadamard_tests}.

Clearly, we only need to optimize the blocks in the causal cone of the Trotter term, see Fig.~\ref{fig:PQCAndCones}~(b). We call this approach the \textsc{cone update}. In \textsc{cone update}, each Hadamard test requires $O(N_b N_p)$ controlled gates where $N_b$ is the number of blocks in the causal cone, and $N_p$ is the number of parameters in a block.

The closer $U$ and $V$ remain during the execution of the algorithm, the fewer controlled gates are required for the Hadamard test. This suggests a systematic way to reduce hardware requirements by introducing approximations to the objective. The first level of approximation consists of replacing $\ket{\psi_{k-1}}$ in Eq.~\eqref{eq:main_objective} with the current variational state $\ket{\psi(\boldsymbol{\theta})}$ after a block has been updated. This way, $U$ and $V$ differ by $O(N_p)$ parameters, greatly simplifying the Hadamard test. Note that the replacement is performed once for each of the $N_b$ blocks in the causal cone. Hence, to effectively simulate time step $\tau$ we use a time step $\tau/N_b$ in the objective.
The division of $\tau$ by $N_b$ is not necessary when $\tau$ is a hyperparameter as, e.g., in imaginary time evolution.
We call this approach the \textsc{block update}.

The second level of approximation consists of replacing $\ket{\psi_{k-1}}$ in the objective with $\ket{\psi(\boldsymbol{\theta})}$ after an angle has been updated. This guarantees that $U$ and $V$ differ by one parameter at all times. Since the replacement is done $N_b N_p$ times, we use a time step $\tau/(N_b N_p)$ in the objective.
It is not necessary to divide $\tau$ by $N_b N_p$ when $\tau$ is a hyperparameter, e.g., in imaginary time evolution.
We call this approach the \textsc{angle update}. Figure~\ref{fig:FHUAndHEHU}~(b) shows an example with five parameters and where the Hadamard test requires a single controlled gate. \textsc{angle update} can be further simplified by replacing the indirect measurement with direct ones~\cite{li2017hybrid,mitarai2019methodology}. This removes the need for the ancillary qubit and the controlled gate resulting in the circuit shown in Fig.~\ref{fig:FHUAndHEHU}~(c).
The coordinatewise update rule for this case are derived in Appendix~\ref{a:hardware-efficient}.

\begin{table*}[t!]
\setlength{\tabcolsep}{7pt}
\renewcommand{\arraystretch}{1.5}
\begin{tabular}{c|c|c|c}
Update method & Circuits per sweep & Matrix inversion & Controlled gates per circuit\\
\hline\hline
\textsc{cone} & $O(N_b N_p)$ & No & $O(N_b N_p)$ \\ 
\hline
\textsc{block} & $O(N_b N_p)$ & No & $O(N_p)$ \\
\hline
\textsc{angle} & $O(N_b N_p)$ & No & $O(1)$ \\
\hline
\textsc{tdvp} & $O( N_b^2 N_p^2 )$ & Yes & $O(1)$ \\
\end{tabular}
\caption{Characteristics of \textsc{cone}, \textsc{block}, and \textsc{angle update} along with the \textsc{tdvp} methods of Ref.~\cite{YuEtAl19}. A sweep consists of updating all parameters inside the causal cone of a Trotter term exactly once.
}
\label{t:updates}
\end{table*}

\section{Results}
\label{sec:Results}

\subsection{Error analysis}
\label{sec:Errors}

In our methods there exist two sources of error.
Firstly, there is a Trotter error resulting from the Trotter product formula that is used to split the time evolution operator. Secondly, there is a variational error due to the limitations of our PQC Ansatz and optimization methods. Both errors can be quantified and systematically decreased.

The Trotter error is determined by the order of the Trotter product formula and can be decreased by using higher-order formulas~\cite{HaSu05}.
A $p$th-order Trotter product formula has an error scaling as $O(\tau^{p+1})$ per time step $\tau$ for sufficiently small values of $\tau$. Then the total Trotter error for the complete time evolution over $N = t/\tau$ time steps scales as $O(\tau^{p})$.

The variational error is determined by the expressive power of the Ansatz and can be decreased by adding new blocks to the PQC and performing more optimization sweeps. A \emph{sweep} consists of updating all parameters inside the causal cone of a Trotter term exactly once. For the $k$th Trotter term, the variational error is simply the objective function in Eq.~\eqref{eq:main_objective}.

\begin{figure}
\centering
\includegraphics[width=.405\textwidth]{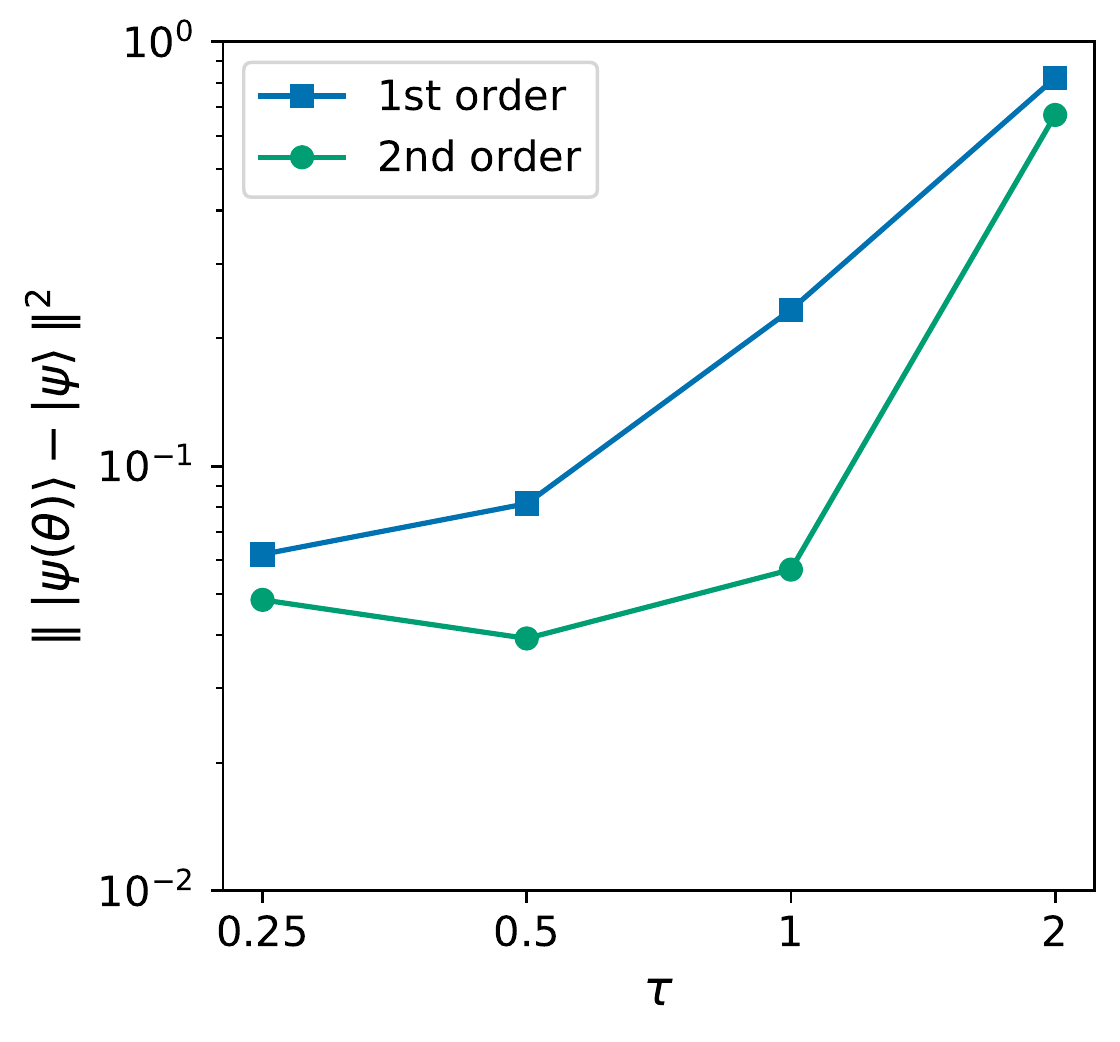}
\vspace{-4mm}
\caption{Error as a function of the time step $\tau$ for hardware-efficient real time evolution with total time $t=2$ using first- and second-order Trotter product formulas.
We observe that for the larger values of $\tau$, i.e., fewer time steps making up the total evolution, the Trotter error is larger than the variational error.
For the smaller values of $\tau$, i.e., more time steps in the entire evolution, the variational error accumulates over more time steps and rapidly becomes larger than the Trotter error.
This observation is consistent with similar studies that were carried out for matrix product states~\cite{Ga06}.
}
\label{fig:trotter_error}
\end{figure}

We obtain numerical evidence that our method can benefit from a second-order Trotter product formula. We perform real time evolution of the $\ket{\boldsymbol{0}}$ state under the quantum Ising chain with $J=1$, $\lambda=0.2$, $n=6$, and $t=2$. In order to achieve a small variational error we use our most accurate algorithm, \textsc{cone update}, with six sweeps. Figure~\ref{fig:trotter_error} shows the squared Euclidean distance between the variational state $\ket{\psi(\boldsymbol{\theta})}$ and the exact evolved state $\ket{\psi}$ as a function of the time step $\tau$ used in the algorithm. The second-order method resulted in smaller errors for all values of $\tau$.

Recently, machine learning heuristics have been employed to assist time evolution algorithms and reduce both the variational and Trotter errors~\cite{zhukov2021quantum,cao2021quantum}.

\begin{figure*}[!ht]
\begin{minipage}[t]{.41\textwidth}
\centering
\includegraphics[width=\linewidth]{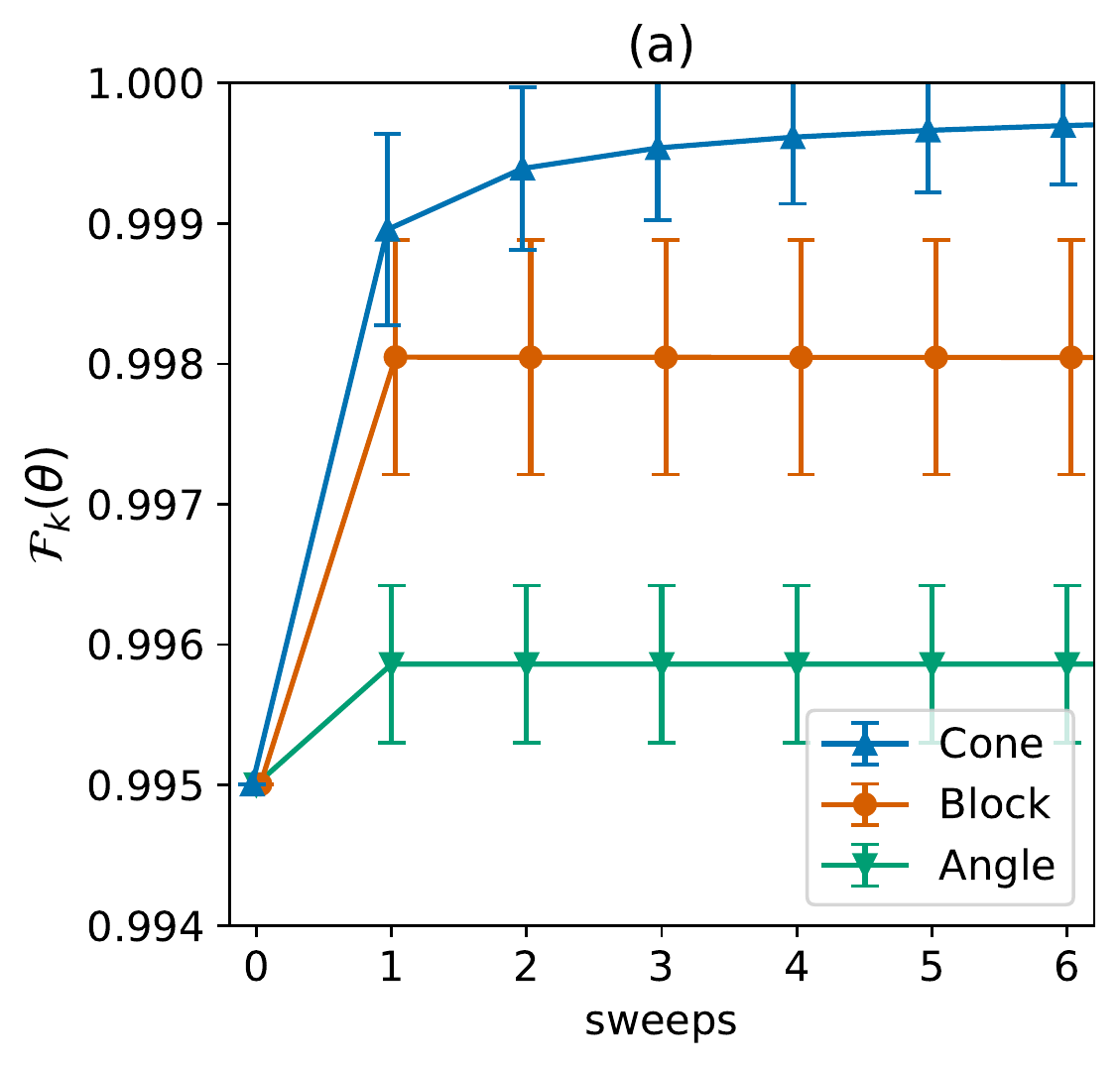}
\end{minipage}
\hspace{.08\textwidth}
\begin{minipage}[t]{.41\textwidth}
\centering
\includegraphics[width=\linewidth]{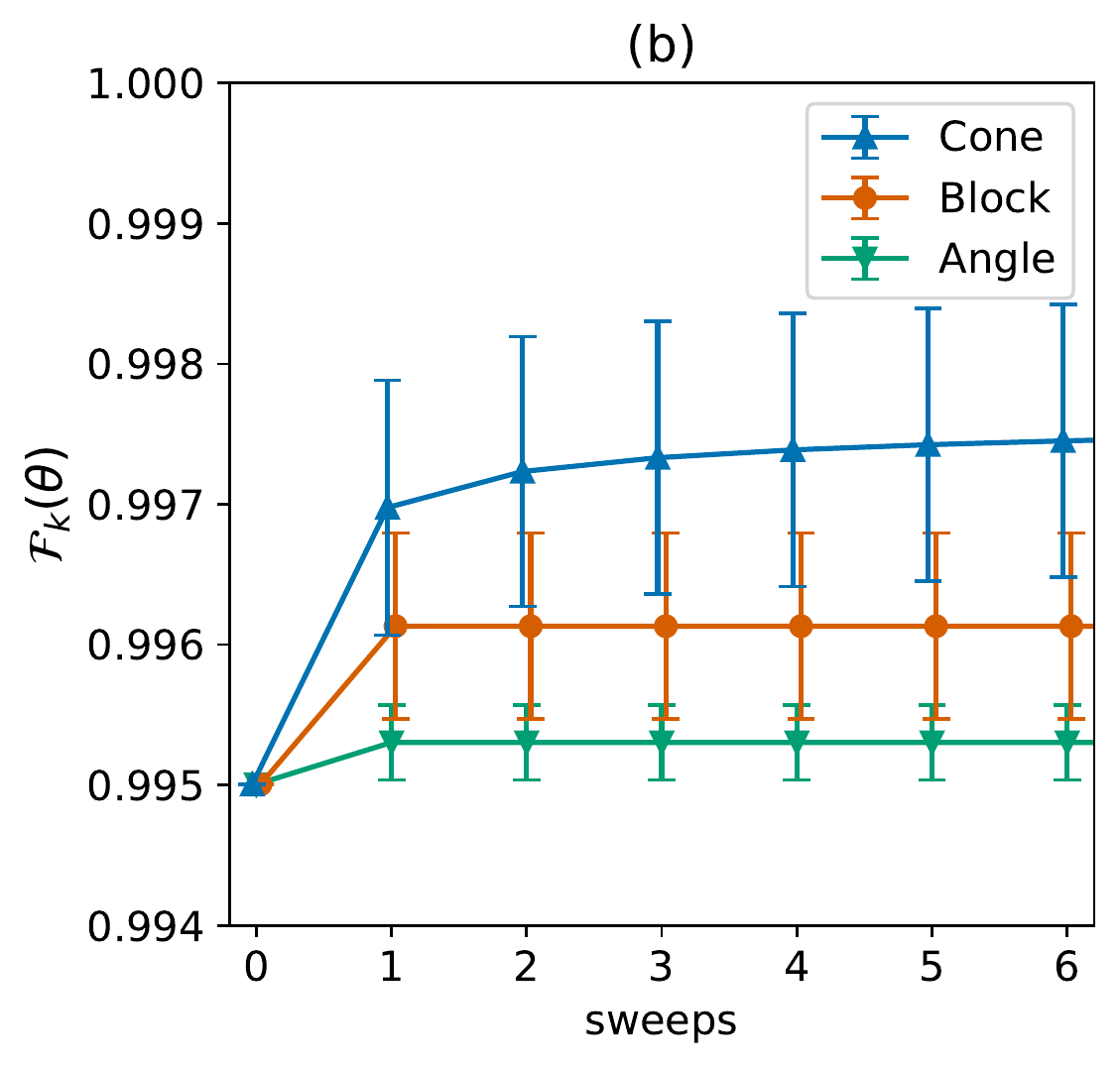}
\end{minipage}
\vspace{-2mm}
\caption{Mean objective function in Eq.~\eqref{eq:main_objective} and standard deviation for $25$ random initial states and Trotter terms of the form $\exp( - \tfrac{i}{10} \hat{\sigma}_{j} \otimes \hat{\sigma}_{j+1} )$, where $\hat{\sigma}$ is chosen randomly from $\{ \mathds{1}, \hat{\sigma}^{x}, \hat{\sigma}^{y}, \hat{\sigma}^{z} \}$. (a) The Trotter term is located in front of a block in our PQC Ansatz. Its causal cone encloses four qubits and three blocks, for a total of $45$ parameters.
(b) The Trotter term is located in front of two blocks in our PQC Ansatz, which is a less favorable position. Its causal cone encloses six qubits and five blocks, for a total of $75$ parameters.
}
\label{fig:experiment1}
\end{figure*}

\begin{figure*}[!ht]
\begin{minipage}[t]{.41\textwidth}
\centering
\includegraphics[width=\linewidth]{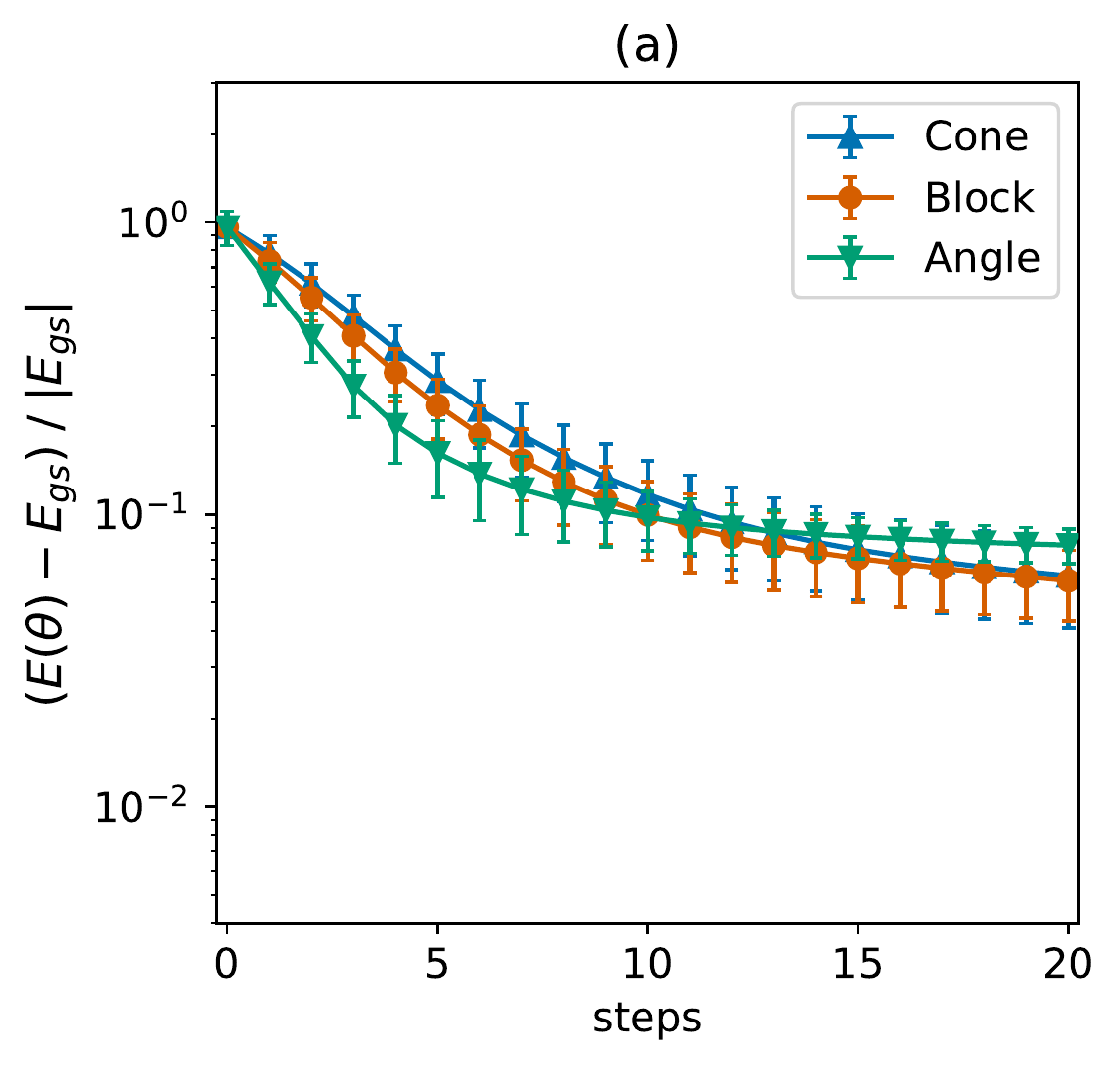}
\end{minipage}
\hspace{.08\textwidth}
\begin{minipage}[t]{.41\textwidth}
\centering
\includegraphics[width=\linewidth]{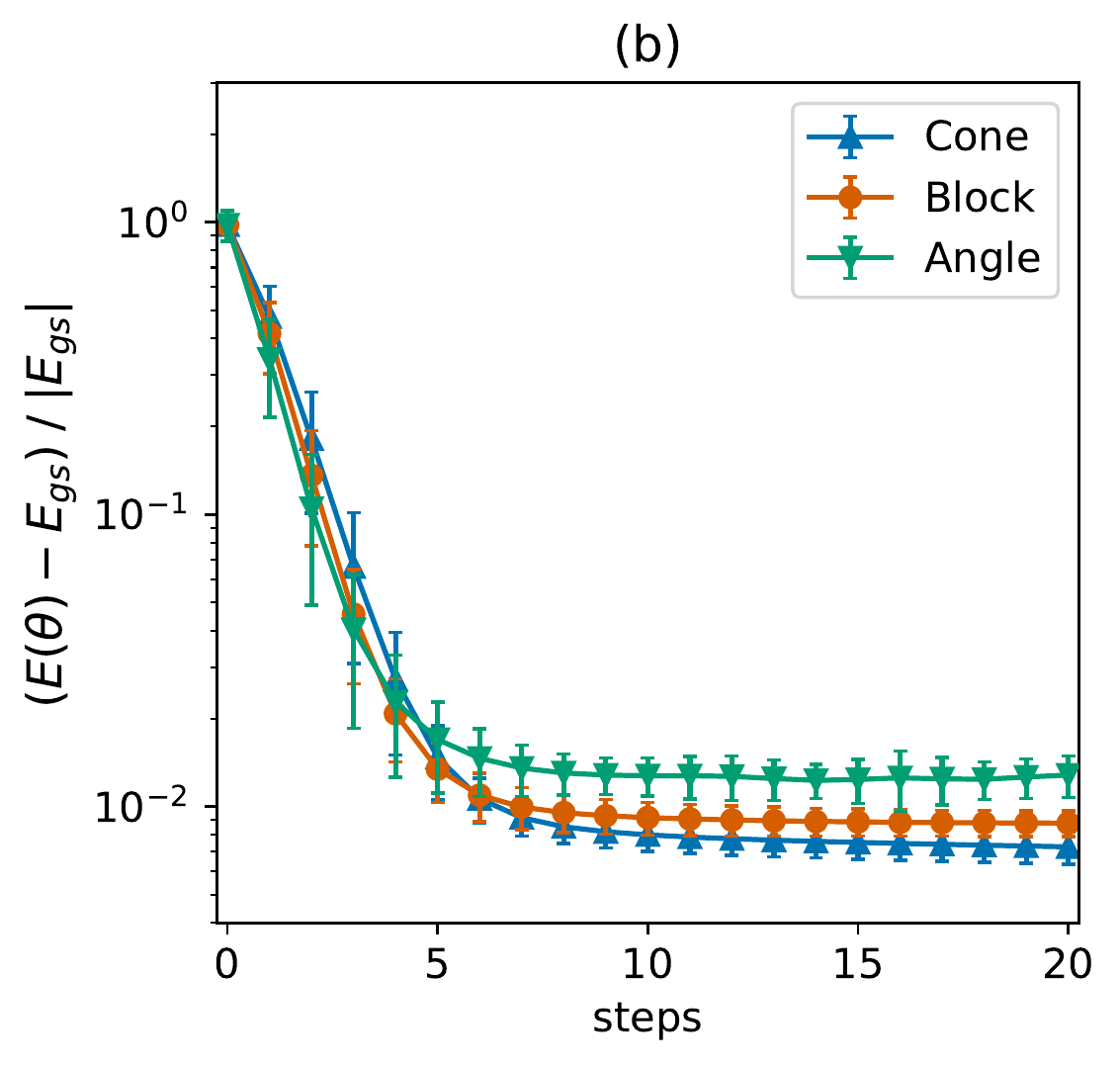}
\end{minipage}
\vspace{-3mm}
\caption{Mean energy and standard deviation of $20$ random initializations for imaginary time evolution on the $1$d quantum Ising model for eight qubits. All the causal cones enclose six qubits at most, thus we are attacking a problem that is larger than the quantum hardware. The $y$ axis is the energy relative to the ground state, where $E(\theta)$ is the variational state energy and $E_{gs}$ is the ground state energy. The time step is a hyperparameter, which we set at $\tau=0.1$ in all cases. (a) The transverse field is set to ${\lambda=1}$ where the corresponding infinite system is critical.
(b) The transverse field is set to ${\lambda=4}$ where the corresponding infinite system is noncritical.}
\label{fig:experiment2}
\end{figure*}

\subsection{Comparison of hardware requirements}

We compare the hardware requirements of our update methods in Table~\ref{t:updates}.
Here $N_b$ denotes the number of blocks inside the causal cone of one Trotter term, $N_p$ denotes the number of parameters per block, and we assume that every block has the same number of parameters.
We conclude that \textsc{angle update} requires the smallest amount of resources per sweep and no matrix inversion which makes this a stable and efficient algorithm for simulating time evolution.
\textsc{block update} interpolates between \textsc{angle} and \textsc{cone update} in a natural way.

Table~\ref{t:updates} also shows the hardware requirements for the \textsc{tdvp} methods of Ref.~\cite{YuEtAl19}.
Compared with these \textsc{tdvp} methods, our update procedures have the advantage that they do not need matrix inversion.
Matrix inversion is numerically unstable when the condition number of the matrix is large and small errors in the matrix can become large errors in the matrix inverse~\cite{GoVa96}.
The \textsc{tdvp} methods of Ref.~\cite{YuEtAl19} compute the matrix elements from mean values over several measurements. For a desired accuracy $\epsilon$ per matrix element, this procedure requires $O(1/\epsilon^{2})$ measurements.
To determine the accuracy of the time-evolved parameters after a \textsc{tdvp} udpate, we need to take into account the condition number $\kappa$ of the matrix because the \textsc{tdvp} update needs the matrix inverse.
We show in Appendix~\ref{a:TDVP} that for a desired accuracy $\epsilon$ in the time-evolved parameters the required number of measurements scales as $O(\kappa^{2}/\epsilon^{2})$ in the worst case.
Therefore, for ill-conditioned matrices, the \textsc{tdvp} methods of Ref.~\cite{YuEtAl19} may need to compute matrix elements very accurately and require many measurements.

We obtain some numerical evidence by computing the median condition number $\tilde{\kappa}$ of $100$ random initializations of our Ansatz. To avoid instabilities, we ignore singular values smaller than $10^{-7}$. With $15$ parameters we have $\tilde{\kappa} \approx 16$, with $45$ parameters we have $\tilde{\kappa} \approx 5001$, and with $75$ parameters we have $\tilde{\kappa} \approx 17085$. This rapid increase of the condition number as a function of the number of variational parameters indicates a challenging scaling of \textsc{tdvp} cost for our choice of Ansatz.

\subsection{Numerical experiments}

We present numerical experiments that validate our methods (details are provided in Appendix~\ref{a:numerical}). First, we look into the accuracy of \textsc{cone}, \textsc{block}, and \textsc{angle update}. This is important, in particular, for real time evolution where the variational state should closely track the true evolved state at all times. In other words, the objective in Eq.~\eqref{eq:main_objective} should be very close to one for any Trotter term. It is interesting to analyze the accuracy as a function of the number of sweeps.

We use the Ansatz in Fig.~\ref{fig:PQCAndCones}~(a) with depth $\mathcal{D}=2$, periodic boundary conditions, and randomly initialized parameters. Regardless of the number of qubits $n$, there are only two possible causal cones: a four-qubit cone enclosing $N_b=3$ blocks, if the term is located in front of a block, and a six-qubit cone enclosing $N_b=5$ blocks, if the term is located in front of two blocks (the latter case is shown in Fig.~\ref{fig:PQCAndCones}~(b)).
We select random Trotter terms of the form $\exp(-\tfrac{i}{10} \hat{\sigma}_{j} \otimes \hat{\sigma}_{j+1})$, where $\hat{\sigma}$ is chosen randomly from $\{ \mathds{1}, \hat{\sigma}^{x}, \hat{\sigma}^{y}, \hat{\sigma}^{z} \}$, and we perform \textsc{cone update} with a time step of $0.1$.
Recall that for real time evolution \textsc{angle} and \textsc{block update} require a corrected time step. For \textsc{block update} we use a step of $0.1/(N_s N_b)$ and for \textsc{angle update} we use $0.1/(N_s N_b N_p)$, where $N_s$ is the number of sweeps.

Figure~\ref{fig:experiment1}~(a) shows the mean objective and standard deviation for the case of four-qubit cones. \textsc{cone update} outperforms and converges to the optimal value of one, while \textsc{block} and \textsc{angle update} do not benefit from an increased number of sweeps. The six-qubit case is shown in Fig.~\ref{fig:experiment1}~(b), where a similar behavior is observed. However, none of the methods converge to the optimal value of one, reflecting the limitations of the shallow PQC Ansatz.

Second, we verify that \textsc{cone}, \textsc{block}, and \textsc{angle update} can find ground states via imaginary time evolution. We use the $1$d quantum Ising Hamiltonian:
\eq{hamiltonian}{
\hat{H}= -J \left( \sum_{j=1}^{n-1} \hat{\sigma}^z_j \hat{\sigma}^z_{j+1} + \lambda \sum_{j=1}^n \hat{\sigma}^x_j \right).
}
For $n=8$ qubits we use the PQC Ansatz in Fig.~\ref{fig:PQCAndCones}~(a) with depth $\mathcal{D}=2$ and open boundary conditions. Here the time step plays the role of a hyperparameter because we are interested in finding the ground state as quickly as possible. We use $\tau = 0.1$ in all cases. 

Figure~\ref{fig:experiment2}~(a) shows the mean energy and standard deviation obtained in $20$ time steps for $J=\lambda=1$ in the Hamiltonian, i.e., for the critical point of the corresponding infinite system. For our choice of hyperparameter $\tau$, all methods reach similar low energies. Figure~\ref{fig:experiment2}~(b) shows the results for $J=1$ and $\lambda=4$, i.e., where the corresponding infinite system is far from the critical point. All methods converge rapidly, producing states that are very close to the ground state. We emphasize that the circuits for \textsc{angle update} are much simpler than those for \textsc{cone update}.

Finally, we test our algorithms on larger instances of up to $12$ qubits. The experiments and results are shown in Fig.~\ref{fig:results} and explained in the figure caption of Fig.~\ref{fig:results}.

\begin{figure}
\centering
\includegraphics[width=75.206mm]{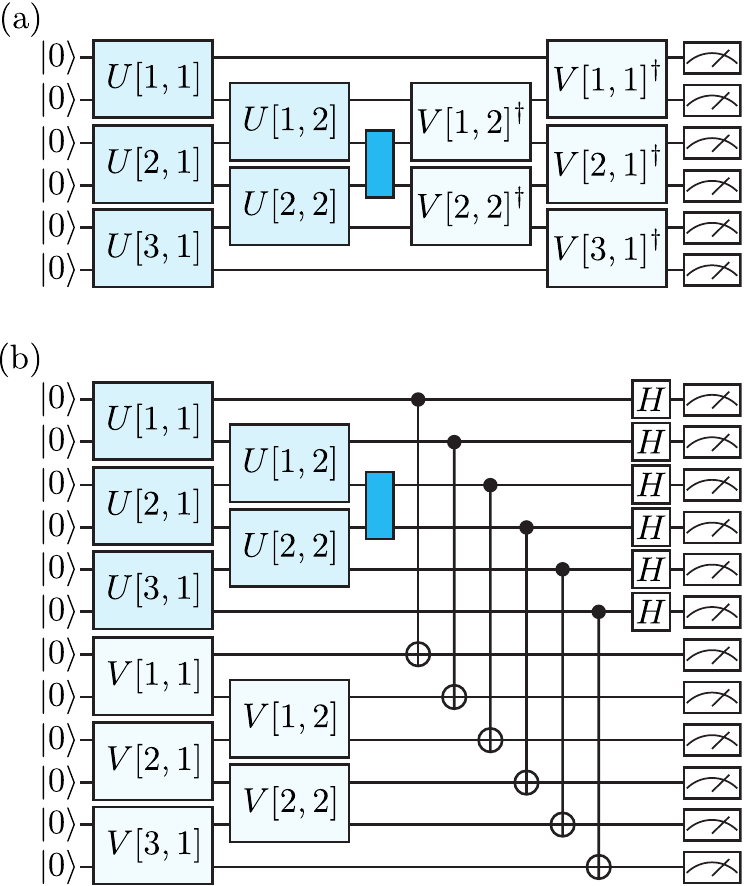}
\caption{Circuits for an alternative hardware-efficient algorithm. Instead of maximizing Eq.~\eqref{eq:main_objective}, this version maximizes the state overlap $\mathcal{F}_k(\boldsymbol{\theta}) = \abs*{\bra{\psi_{k-1}} e^{i \tau h_k \hat{H}_k} \ket{\psi(\boldsymbol{\theta})}}^{2}$, where $\ket{\psi_{k-1}} = V \ket{\boldsymbol{0}}$ and $\ket{\psi(\boldsymbol{\theta})}=U(\boldsymbol{\theta})\ket{\boldsymbol{0}}$. (a) The probability of measuring $\ket{\boldsymbol{0}}$ is equivalent to the state overlap. (b) The state overlap can alternatively be computed using this shorter-depth circuit representing the destructive Swap test~\cite{GaCh13, CiEtAl18}.
}
\label{fig:HECU}
\end{figure}

\section{Discussion}
\label{sec:Discussion}

Variational simulations of time evolution can be performed without inverting a possibly ill-conditioned matrix at each time step. To this end, we derived suitable algorithms whose hardware requirements can be adjusted to match the experimental capabilities.

One of the main applications of imaginary time evolution is to find ground states. Our most efficient algorithm, \textsc{angle update}, performed remarkably well at this task. In practice, once \textsc{angle update} converges one could switch to more demanding algorithms, such as \textsc{block} or \textsc{cone update}, in order to fine-tune the result.

For real time evolution, the task is to simulate the time-dependent Schr\"{o}dinger equation.
We presented numerical evidence that \textsc{block} and \textsc{cone update} achieve the high accuracy required.
We also expect our \textsc{angle update} to be useful for specific applications, such as the computation of steady states, where accuracy per time step is not crucial.

A recent publication~\cite{JaEtAl20} contains simulations of real time evolution based on the maximization of the state overlap.
As illustrated in Fig.~\ref{fig:HECU}, combining this method with our cone strategy leads to a promising hardware-efficient algorithm that can additionally make use of hardware-efficient overlap computation~\cite{GaCh13, CiEtAl18}. Here the coordinatewise update rules are already known~\cite{ViTh18, NaFuTo19, PaIoOzMc19, ostaszewski2019quantum} as the overlap maximization is equivalent to the expectation minimization for the Hermitian operator $M=-\dyad{\boldsymbol{0}}$.

A number of recent papers implement tensor network techniques via parameterized quantum circuits (PQCs), see Refs.~\cite{KiSw17, grant2018hierarchical,huggins2019towards,LuEtAl20, smith2020crossing, BarEtAl20, LiEtAl20, MaEtAl21} for examples. Our work contributes to this line of research, bringing PQC and tensor network optimization closer together.

The last key aspect of this work is the use of causal cones to enable simulations of finite systems larger than the size of the underlying quantum hardware. Causal cones are also an ingredient in the construction of noise-resilient quantum circuits~\cite{KiSw17, Ki17, BoChFr19}. We envision that such techniques that detach the logical model from some of the hardware limitations will ultimately enable us to attack large problems and obtain a quantum advantage.

\section{Acknowledgements}
We thank Nathan Fitzpatrick, David Amaro, and Carlo Modica for helpful discussions. We gratefully acknowledge the cloud computing resources received from the `Microsoft for Startups' program.

\onecolumngrid
\appendix

\section{Variational simulation of time evolution}
\label{a:variational}

In this Appendix, we detail our method for the variational simulation of time evolution. To keep the discussion general, we consider an arbitrary complex time $z \in \mathbb{C}$. Later we will specialize to purely real time and purely imaginary time evolution.

We want to simulate the time evolution operator $e^{-i z \hat{H}}$ applied to an initial state $\ket{\psi}$ of $n$ qubits. We assume that the Hamiltonian is given in general form $\hat{H} = \sum_{k=1}^K h_k \hat{H}_k$, where $\hat{H}_k \in \{\mathds{1}, \hat{Z}, \hat{X}, \hat{Y}\}^{\otimes n}$ is a tensor product of Pauli operators, $h_k$ is a real number, and $K \sim O(\text{poly}(n))$. There can be terms $\hat{H}_{k}$ in $\hat{H}$ that do not commute with each other.

A commonly used technique to simplify the problem consists of expanding the time evolution operator into a product. A product formula, such as the Trotter formula, produces a sequence of short-time evolutions which approximates the full evolution. Let us apply the first-order Trotter product formula for $N \sim O(\text{poly}(n))$ time steps of size $\zeta=z/N$:
\eq{variational1}{
e^{-i z \hat{H}}\ket{\psi} &\approx \left( e^{-i\zeta h_K \hat{H}_K} \cdots e^{-i\zeta h_1 \hat{H}_1} \right)^N \ket{\psi}\\
&= e^{-i\zeta h_K \hat{H}_{K,N}} \cdots e^{-i\zeta h_1 \hat{H}_{1,N}} \cdots e^{-i\zeta h_K \hat{H}_{K,1}} \cdots e^{-i\zeta h_1 \hat{H}_{1,1}} \ket{\psi} .
}
In the second line, $\hat{H}_{k,m}$ has an additional subscript indicating the $m$th application of the $k$th term. Now assume we are able to obtain a variational approximation to the first operation:
\eq{variational2}{
e^{-i\zeta h_1 \hat{H}_{1,1}} \ket{\psi} \approx \ket{ \psi(\boldsymbol{\theta}_{1,1}^*)} ,
}
where $\boldsymbol{\theta}_{1,1}$ is the vector of variational parameters and $\boldsymbol{\theta}_{1,1}^*$ indicates their optimal value. Then we can substitute this in Eq.~\eqref{eq:variational1} and proceed with a variational approximation to the second operation:
\eq{variational3}{
e^{-i \zeta h_2 \hat{H}_{2,1}} e^{-i \zeta h_1 \hat{H}_{1,1}} \ket{\psi} &\approx e^{-i \zeta h_2 \hat{H}_{2,1}} \ket{ \psi(\boldsymbol{\theta}_{1,1}^*)} \approx \ket{ \psi(\boldsymbol{\theta}_{2,1}^*)} .
}
where again we assumed the optimal value for the variational parameters. Iterating the above procedure for all the $NK$ terms we obtain an approximation to the full time evolution $e^{-i z \hat{H}} \ket{\psi} \approx \ket{ \psi(\boldsymbol{\theta}_{K,N}^*)}$. To simplify the notation let us condense indexes $k$ and $n$ into a single index $l$, and let us write $\ket{\psi(\boldsymbol{\theta}^*_l)} = \ket{\psi_l}$ whenever the parameters have been optimized.

In order to find the variational approximation at step $l$ we use the squared Euclidean distance as a cost function:
\eq{variational4}{
\mathcal{C}_l(\boldsymbol{\theta}_l) &= \norm{\ket{\psi(\boldsymbol{\theta}_l)} - e^{-i \zeta h_l \hat{H}_l} \ket{\psi_{l-1}} }^{2}\\
&= \braket{\psi(\boldsymbol{\theta}_l)}{\psi(\boldsymbol{\theta}_l)} + \bra{\psi_{l-1}} e^{i(\bar{\zeta} - \zeta) h_l \hat{H}_l } \ket{\psi_{l-1}} - \bra{\psi(\boldsymbol{\theta}_l)} e^{-i \zeta h_l \hat{H}_l} \ket{\psi_{l-1}} - \bra{\psi_{l-1}} e^{i \bar{\zeta} h_l \hat{H}_l} \ket{\psi(\boldsymbol{\theta}_{l})}\\
&= \text{const.} - 2 \Re \left( \bra{\psi_{l-1}} e^{i \bar{\zeta} h_l \hat{H}_l} \ket{\psi(\boldsymbol{\theta}_{l})}
\right) .}
Here $\Re(\cdot)$ denotes the real part of a complex number, and $\bar{\zeta}$ denotes the complex conjugate of $\zeta$. We have assumed an Ansatz for which $\braket{ \psi(\boldsymbol{\theta}_l)}{ \psi(\boldsymbol{\theta}_l)}$ is constant. This is the case, for example, if the Ansatz is implemented by a PQC. We have also used that $\hat{H}_l$ is Hermitian.

The minimization of $\mathcal{C}_l(\boldsymbol{\theta}_l)$ is equivalent to the maximization of the following objective function:
\eq{variational_obj}{
\mathcal{F}_l(\boldsymbol{\theta}_l) &= \Re \left( \bra{\psi_{l-1}} e^{i \bar{\zeta} h_l \hat{H}_l} \ket{\psi(\boldsymbol{\theta}_{l})} \right).
}

Let us now specialize to the two cases of interest. For real time evolution we write $z \equiv t$ where $t \in \mathbb{R}$ is the total time, and $\zeta \equiv \tau = t/N$ is the time step. Since the terms $\hat{H}_{l}$ are tensor products of Pauli operators, we have that $\hat{H}_{l}^2 = \mathds{1}$. Using this property and the definition of matrix exponential $e^A = \sum_{n=0}^\infty A^n / n!$, it can be verified that $e^{i \tau h_l H_l} = \cos(\tau h_l) \mathds{1} + i\sin(\tau h_l) \hat{H}_l$. Plugging this in the objective function, we obtain:
\eq{variational_obj_real}{
\mathcal{F}_{l,\text{real}}(\boldsymbol{\theta}_l) &= \cos(\tau h_l) \Re \Big( \braket{\psi_{l-1}}{\psi(\boldsymbol{\theta}_{l})} \Big) - \sin(\tau h_l) \Im \left( \bra{\psi_{l-1}} \hat{H_l} \ket{\psi(\boldsymbol{\theta}_{l})} \right) .
}

For imaginary time, we write $z \equiv -i t$, where $t \in \mathbb{R}$ is the total time, and $\zeta \equiv -i\tau = -i t/N$ is the time step. Following the same argument above, it can be verified that $e^{- \tau h_l H_l} = \cosh(\tau h_l) \mathds{1} - \sinh(\tau h_l) \hat{H}_l$. Plugging this in the objective function we obtain:
\eq{variational_obj_imag}{
\mathcal{F}_{l,\text{imag}}(\boldsymbol{\theta}_l) &= \cosh(\tau h_l) \Re \Big( \braket{\psi_{l-1}}{\psi(\boldsymbol{\theta}_{l})} \Big) - \sinh(\tau h_l) \Re \left( \bra{\psi_{l-1}} \hat{H_l} \ket{\psi(\boldsymbol{\theta}_{l})} \right) .
}

In Appendix~\ref{a:sinusoidal}, we show that the coordinatewise version of Eq.~\eqref{eq:variational_obj} has a sinusoidal form. This fact is inherited by Eqs.~\eqref{eq:variational_obj_real} and~\eqref{eq:variational_obj_imag}, and is exploited to design the optimization algorithm in Appendix~\ref{a:rotosolve}. In Appendix~\ref{a:hadamard_tests}, we present quantum circuits for the estimation of the objectives.

\section{Sinusoidal form of the coordinatewise objective function}
\label{a:sinusoidal}

In Refs.~\cite{ViTh18, NaFuTo19, PaIoOzMc19, ostaszewski2019quantum}, the authors showed that for certain standard PQCs the expectation $\tr(M U \rho U^\dag)$ as a function of a single parameter has sinusoidal form. This yields an efficient coordinatewise optimization algorithm that does not require explicit computation of the gradient or the Hessian. Here we present a similar derivation for objective functions of the form $\Re( \tr( M U \rho V^\dag ) )$ where $U$ is a PQC and $V$ is a fixed circuit. Note that there is nothing preventing us from parametrizing $V$ and carrying out the same derivation.

Let us consider a circuit of the form $U(\boldsymbol{\theta}) = U_D \cdots U_1$, where each gate is either fixed, e.g., a CNOT, or parameterized as $U_d = \exp(-i \theta_d G_d)$, where $\theta_d \in (-\pi, \pi]$ and $G_d$ is a Hermitian and unitary matrix such that $G_d^2=\mathds{1}$. For example, tensor products of Pauli matrices are suitable choices for ${G_d \in \{ \mathds{1}, \hat{X}, \hat{Y}, \hat{Z}\}^{\otimes n}}$. For the parameterized gates, we use the definition of matrix exponential to get $U_d = \cos ( \theta_d ) \mathds{1} - i\sin (\theta_d) G_d$. Without loss of generality, let us consider a pure initial state $\rho = \dyad{\mathbf{0}}$ where $\ket{\mathbf{0}} \equiv \ket{0}^{\otimes n}$.

Expanding the objective function we have $\Re( \bra{\mathbf{0}} V^\dag M U_D \cdots U_d \cdots U_1 \ket{\mathbf{0}} )$. To express this as a function of a single parameter $\theta_d$ we simplify the notation absorbing all gates before $U_d$ in a unitary which we call $U_B$, and we absorb all gates after $U_d$ in a unitary which we call $U_A$. Using this notation, we write:
\eq{exp_1}{
f_d(x)
&= \Re \left( \bra{\mathbf{0}} V^\dag M U_A U_d U_B \ket{\mathbf{0}} \right) \\
&= \Re \left( \bra{\mathbf{0}} V^\dag M U_A \left( \cos ( x ) I - i \sin ( x ) G_d \right) U_B \ket{\mathbf{0}} \right) \\
&= \Re \left( \bra{\mathbf{0}} V^\dag M U_A U_B \ket{\mathbf{0}} \right) \cos ( x ) + \Re \left( \bra{\mathbf{0}} V^\dag M U_A (- i G_d) U_B \ket{\mathbf{0}} \right) \sin (x) .
}
Noting that $U_d(0) = I$ and $U_d(\tfrac{\pi}{2}) = -i G_d$, we rewrite the above as:
\eq{exp_2}{
f_d(x) &= \Re \left( \bra{\mathbf{0}} V^\dag M U_A U_d(0) U_B \ket{\mathbf{0}} \right) \cos ( x ) + \Re \left( \bra{\mathbf{0}} V^\dag M U_A U_d(\tfrac{\pi}{2} ) U_B \ket{\mathbf{0}} \right) \sin ( x ) \\
&= f(0) \cos ( x ) + f(\tfrac{\pi}{2}) \sin ( x ) .
}
We now use the harmonic addition theorem $a \cos(x) + b \sin(x) = \sqrt{a^2 + b^2} \sin(x + \arctan(\frac{a}{b}))$ to obtain: 
\eq{exp_3}{ 
f_d(x) &= A \sin( x + B) , \\
A &= \sqrt{ f_d(0)^2 + f_d(\tfrac{\pi}{2})^2 }, \\
B &= \arctantwo \left( f_d(0), \; f_d(\tfrac{\pi}{2}) \right ) .
}
The objective function as a function of a single parameter has sinusoidal form with amplitude $A$, phase $B$, and period $2\pi$. Note that we must use the \texttt{arctan2} function in order to correctly handle the sign of numerator and denominator, as well as the case where the denominator is zero.

There is nothing special about the evaluations at $0$ and $\tfrac{\pi}{2}$ in Eq.~\eqref{eq:exp_3}. Indeed, we can estimate $A$ and $B$ from:
\eq{general_estimator_A}{
\sqrt{ f_d(\phi)^2 + f_d(\phi + \tfrac{\pi}{2})^2 } = \sqrt{ A^2 \sin^2(\phi +B) + A^2 \sin^2(\phi + \tfrac{\pi}{2} + B ) } = \abs{A} \sqrt{ \sin^2(\phi +B) + \cos^2(\phi + B ) } = A ,
}
\eq{general_estimator_B}{
\frac{f_d(\phi)}{f_d(\phi + \tfrac{\pi}{2})} = \frac{\sin(\phi + B )}{\sin(\phi + \tfrac{\pi}{2} + B )} = \tan(\phi + B) .
}
for any $\phi \in \mathbb{R}$. 

From the graph of the sine function, it is easy to locate the maxima at ${ \theta_d^* = \tfrac{\pi}{2} - B + 2\pi k}$ for all $k \in \mathbb{Z}$. Taking $B$ from Eq.~\eqref{eq:general_estimator_B}, we obtain:
\eq{real_min_phi}{
\theta_d^* &= \argmax_{x} f_d(x) \\
&= \tfrac{\pi}{2} - \arctantwo \left( f_d(\phi) , \; f_d(\phi + \tfrac{\pi}{2}) \right ) +\phi +2\pi k .
}
In practice, we choose $k$ such that $\theta_d^* \in (-\pi,\pi]$.

\begin{figure}
\centering
\includegraphics[width=.41\textwidth]{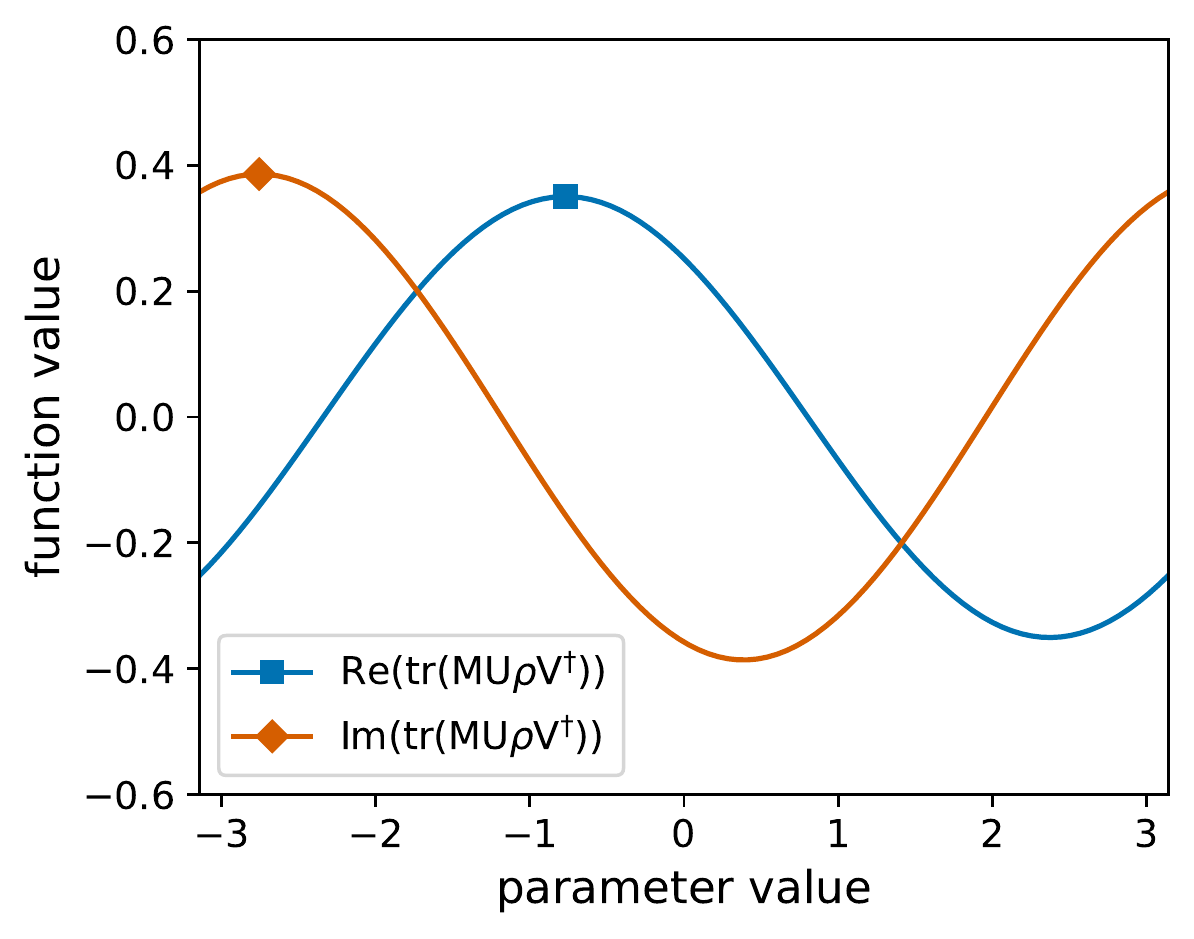}
\vspace{-3mm}
\caption{Sinusoidal form of the objective functions used in this work. This example is for a random choice of $U, V$, $M$, and $\rho$ on $n=4$ qubits. The maxima (square and diamond) can be found in closed form. This requires the evaluation of the objective function at two arbitrary parameter values spaced $\tfrac{\pi}{2}$ apart.}
\label{f:sinusoidal}
\end{figure}

A similar derivation can be done for objective functions of the form $\Im(\tr(M U \rho V^\dag))$. For the $d$th parameter, we write $f_d(x) = \Im( \bra{\mathbf{0}} V^\dag M U_A U_d U_B \ket{\mathbf{0}} )$ and we obtain the maxima exactly as in Eq.~\eqref{eq:real_min_phi}. Figure~\ref{f:sinusoidal} shows the sinusoidal forms for a random choice of $U, V$ and $M$ on $n=4$ qubits.

\section{The update rule with a single evaluation}
\label{a:rotosolve}

In this Appendix, we take a closer look at the quantity of interest, Eq.~\eqref{eq:variational_obj}, and derive our coordinatewise update rule. Consider the $l$th term in the Trotter product formula. The coordinatewise objective for the $d$th parameter is sinusoidal:
\eq{rotosolve1}{
f_{l,d} (x) &= \mathcal{F}_l(\theta_1, \cdots, \theta_{d-1}, x, \theta_{d+1}, \cdots, \theta_D)\\
&= \Re \left( \bra{\mathbf{0}} V^\dag e^{i \bar{\zeta} h_l \hat{H}_l} U_D \cdots U_d(x) \cdots U_1 \ket{\mathbf{0}} \right)\\
&= \underbrace{\sqrt{ f_{l,d}(\phi)^2 + f_{l,d}(\phi + \tfrac{\pi}{2})^2 }}_{A_{l,d}} \sin \Big( x + \underbrace{\arctantwo \left( f_{l,d}(\phi) , \; f_{l,d}(\phi + \tfrac{\pi}{2}) \right) -\phi}_{B_{l,d}} \Big),\\
}
where $A_{l,d}$ is the amplitude, $B_{l,d}$ is the phase, and $\phi \in \mathbb{R}$ can be chosen at will. The third line is obtained by applying the results in Eqs.~\eqref{eq:exp_3},~\eqref{eq:general_estimator_A} and~\eqref{eq:general_estimator_B}.

Optimizing this objective using Eq.~\eqref{eq:real_min_phi} would require evaluations of $f_{l,d}(\phi)$ and $f_{l,d}(\phi+\tfrac{\pi}{2})$. However, we can recycle information from previous steps and use a single evaluation. The approach is as follows. Say we have found the maximum $\theta_{d-1}^*$ for the $(d-1)$th parameter. At no additional cost, we calculate $f_{l,d-1}(\theta_{d-1}^*) = A_{l,d-1}$. Now we move to the $d$th parameter. Setting $\phi$ in Eq.~\eqref{eq:rotosolve1} to the current parameter value, $\phi = \theta_{d}$, we happen to know $f_{l,d}(\phi) = f_{l,d}(\theta_{d}) = f_{l,d-1}(\theta_{d-1}^*) = A_{l,d-1}$. Hence, we only need to evaluate $f_{l,d}(\phi + \tfrac{\pi}{2}) = f_{l,d}(\theta_{d} + \tfrac{\pi}{2})$.

In summary, we obtain the update rule: 
\eq{rotosolve2}{
\theta_d^* &= \argmax_x f_{l,d} (x)\\
&= \frac{\pi}{2} - \arctantwo \left( f_{l,d}(\theta_{d}), \; f_{l,d}(\theta_d + \tfrac{\pi}{2}) \right) + \theta_d ,
}
where $\theta_d$ is the current parameter value, and $f_{l,d}(\theta_{d})$ is known from the previous parameter update. Equation~\eqref{eq:rotosolve2} is our coordinatewise update rule for the variational simulation of time evolution.

\section{Hadamard test for \textsc{cone update}}
\label{a:hadamard_tests}

In this appendix, we discuss the Hadamard test used in \textsc{cone update}. Recall that the proposed method initializes a PQC at each step and trains it to simulate the action of a short-time evolution operator on the previous variational state. Let us denote the state obtained at the $(l-1)$th step as $\ket{\psi_{l-1}} = V \ket{\mathbf{0}}$ and the state for the $l$th step as $\ket{\psi(\boldsymbol{\theta}_l)} = U \ket{\mathbf{0}}$. Here $V$ and $U$ are PQCs acting on the easy-to-prepare state $\ket{\mathbf{0}} \equiv \ket{0}^{\otimes n}$.

The Hadamard test can be challenging to execute on existing hardware when $U$ and $V$ are unrelated quantum circuits due to the potentially large number of controlled operation. \textsc{cone update} uses the fact that $U$ and $V$ differ only locally to simplify the Hadamard test. One circuit can be efficiently transformed into the other using adjoints of gates. 
To show this we start from $U\ket{\mathbf{0}}$, add one ancilla qubit in the state $\ket{0}$ which is then acted upon by a Hadamard gate, so that we get $\tfrac{1}{\sqrt{2}}(\ket{0}+\ket{1}) \otimes U \ket{\mathbf{0}}$. Now we include the local transformations from $U$ to $V$ as gates controlled by the ancilla qubit. As an example, if $U$ contains a rotation gate $R_z(a)$ and $V$ contains $R_z(b)$ at the same location, then we only need to attach a controlled-$R_z(b-a)$ rotation. The subspaces where the ancilla is $\ket{0}$ and $\ket{1}$ will contain $U \ket{\mathbf{0}}$ and $V \ket{\mathbf{0}}$, respectively. In other words, the result is $\tfrac{1}{\sqrt{2}}(\ket{0} \otimes U \ket{\mathbf{0}} + \ket{1} \otimes V \ket{\mathbf{0}})$. Having another Hadamard gate acting upon the ancilla qubit gives $\tfrac{1}{2} \left( \ket{0} \otimes (U +V) \ket{\mathbf{0}} + \ket{1} \otimes (U-V)\ket{\mathbf{0}} \right)$. Measuring the expectation of $\hat{Z} \otimes \hat{H_l}$ one obtains the real part $\expval*{ \hat{Z} \otimes \hat{H_l} } = \Re (\bra{ \mathbf{0}} V^{\dag} \hat{H_l} U \ket{\mathbf{0}} )$. 

For the imaginary part, we follow the same procedure, but we include a phase gate after the first Hadamard gate. This yields $\expval*{ \hat{Z} \otimes \hat{H_l} } = \Im (\bra{ \mathbf{0}} V^{\dag} \hat{H_l} U \ket{\mathbf{0}} )$. With the two circuits just described we can estimate the objective function in Eq.~\eqref{eq:variational_obj_real} for real time evolution, and in Eq.~\eqref{eq:variational_obj_imag} for imaginary time evolution. If the causal cone of $\hat{H_l}$ contains $N_b$ blocks, each with $N_p$ parameterized gates, the Hadamard test requires $O(N_b N_p)$ controlled operations.

\section{Hadamard test for \textsc{angle update}}
\label{a:hardware-efficient}

In this appendix, we present the implementation of our method that has the lowest hardware requirements. \textsc{angle update} is efficient in terms of circuit depth, but still uses an ancilla qubit and a controlled operation which may be challenging to realize on existing hardware. We can avoid the use of those while requiring the execution of additional circuits. This approach uses the methods presented in Refs.~\cite{li2017hybrid} and~\cite{mitarai2019methodology} to replace indirect measurements with direct ones.

Recall that \textsc{angle update} consists of replacing the variational state $\ket{\psi_{l-1}} = V \ket{\mathbf{0}}$ in the objective function with the variational state $\ket{\psi(\boldsymbol{\theta})} = U \ket{\mathbf{0}}$ after each parameter update. This approximation guarantees that the two PQCs $V$ and $U$ differ by just one parameter at all times. Thus we can drop $V$ and express everything in terms of $U$. 

For real time evolution ($\zeta \equiv \tau \in \mathbb{R}$), we start from Eq.~\eqref{eq:variational_obj_real}. The coordinatewise objective for the $d$th parameter is:
\eq{hardware_efficient2}{
f_{l,d,\text{real}}(x) = &\cos(\tau h_l) \Re \left( \bra{\mathbf{0}}U_1^\dag \cdots U_d^\dag \cdots U_D^\dag \mathds{1} U_D \cdots U_d(x) \cdots U_1 \ket{\mathbf{0}} \right)\\
&- \sin(\tau h_l) \Im \left( \bra{\mathbf{0}} U_1^\dag \cdots U_d^\dag \cdots U_D^\dag \hat{H_l} U_D \cdots U_d(x) \cdots U_1 \ket{\mathbf{0}} \right) .
}
The variable $x$ appears only in the gate denoted by $U_d(x)$, while $U_d^\dag$ is fixed and uses the current parameter value $\theta_d$.

We maximize this objective using Eq.~\eqref{eq:rotosolve2}. The first evaluation is at the current parameter value which yields $f_{l,d,\text{real}}(\theta_d) = \cos(\tau h_l)$. The second evaluation is at the shifted parameter value and is slightly more involved. Using $U_d(\theta_d +\tfrac{\pi}{2}) = U_d(\theta_d) U_d(\tfrac{\pi}{2}) = -i U_d(\theta_d)G_d $, we have that:
\eq{hardware_efficient3}{
f_{l,d,\text{real}}(\theta_d + \tfrac{\pi}{2}) = -\sin(\tau h_l) \Im \left( \bra{\mathbf{0}} U_1^\dag \cdots U_d^\dag \cdots U_D^\dag \hat{H_l} U_D \cdots U_d (-iG_d) \cdots U_1 \ket{\mathbf{0}} \right) .
}

Now we can use the technique from Ref.~\cite{mitarai2019methodology} to write the above as the difference of two expectations. Since $G_d$ is a Pauli operator, we can define projective measurement operators $\tfrac{1}{\sqrt{2}}(\mathds{1} \pm G_d)$. In turn these can be used to define new observables $\hat{H}_{l\pm} = \tfrac{1}{2}(\mathds{1} \pm G_d)U_d^\dag \cdots U_D^\dag \hat{H_l} U_D \cdots U_d (\mathds{1} \pm G_d)$ which are easily verified to be Hermitian. With these, a direct calculation shows that:
\eq{hardware_efficient4}{
f_{l,d,\text{real}}(\theta + \tfrac{\pi}{2}) = -\frac{\sin(\tau h_l)}{2} \left( \expval*{\hat{H}_{l+}} -\expval*{\hat{H}_{l-}} \right) .
}
In practice, the projective measurement operators $\tfrac{1}{\sqrt{2}}(\mathds{1} \pm G_d)$ correspond to measuring the qubit when these operators act in the eigenbasis of $G_{d}$~\cite{mitarai2019methodology}.
Putting everything together, Eq.~\eqref{eq:hardware_efficient2} is maximized in closed form using two expectations:
\eq{hardware_efficient5}{
\theta_d^* &= \argmax_x f_{l,d,\text{real}} (x)\\
&= \frac{\pi}{2} - \arctantwo \left( 2\cot(\tau h_l) , \; \expval*{\hat{H}_{l-}} - \expval*{\hat{H}_{l+}} \right) + \theta_d .
}
All quantities are estimated by direct measurements, without using ancilla qubits or controlled gates.

For imaginary time evolution ($\zeta \equiv -i \tau$ with $\tau \in \mathbb{R}$), we start from Eq.~\eqref{eq:variational_obj_imag}. The coordinatewise objective for the $d$th parameter is:
\eq{hardware_efficient6}{
f_{l,d,\text{imag}}(x) = &\cosh(\tau h_l) \Re \left( \bra{\mathbf{0}}U_1^\dag \cdots U_d^\dag \cdots U_D^\dag \mathds{1} U_D \cdots U_d(x) \cdots U_1 \ket{\mathbf{0}} \right)\\
&- \sinh(\tau h_l) \Re \left( \bra{\mathbf{0}} U_1^\dag \cdots U_d^\dag \cdots U_D^\dag \hat{H_l} U_D \cdots U_d(x) \cdots U_1 \ket{\mathbf{0}} \right) .
}

Again we maximize this objective using Eq.~\eqref{eq:rotosolve2}.
The first evaluation is at the current parameter value which yields $f_{l,d,\text{imag}}(\theta_d) = \cosh(\tau h_l) -\sinh(\tau h_l) \expval*{\hat{H}}_{\theta_d}$. The subscript is used to stress that the expectation is computed using the current parameter value. The second evaluation is at the shifted parameter value. Using $U_d(\theta_d +\tfrac{\pi}{2}) = -i U_d(\theta_d)G_d $ in Eq.~\eqref{eq:hardware_efficient6}, we see that the first term equals to zero. For the second term, we use the technique from Ref.~\cite{li2017hybrid} to express it as the difference of two expectations. The result is:
\eq{hardware_efficient7}{
f_{l,d,\text{imag}} (\theta_d + \tfrac{\pi}{2}) &= -\frac{\sinh(\tau h_l)}{2} \left( \expval*{\hat{H}_l}_{\theta_d + \tfrac{\pi}{4}} - \expval*{\hat{H}_l}_{\theta_d - \tfrac{\pi}{4}} \right) .
}
Here the subscripts are used to stress the use of a shifted value for the $d$th parameter. Putting everything together, Eq.~\eqref{eq:hardware_efficient6} is maximized in closed form using three expectations:
\eq{hardware_efficient8}{
\theta_d^* &= \argmax_x f_{l,d,\text{imag}} (x)\\
&= \frac{\pi}{2} - \arctantwo \left( 2\coth(\tau h_l) - 2\expval*{\hat{H}_l}_{\theta_d} , \; \expval*{\hat{H}_l}_{\theta_d - \tfrac{\pi}{4}} - \expval*{\hat{H}_l}_{\theta_d + \tfrac{\pi}{4}} \right) + \theta_d .
}
The three expectations are estimated by direct measurements, without using ancilla qubits or controlled gates. 

Recall that in \textsc{cone update}, we are able to recycle information from previous steps and reduce the circuit count. In \textsc{angle update}, this cannot be done exactly as the objective function changes after each parameter update. For imaginary time, assuming the change in objective function value is small, we can approximately recycle information from the previous steps as $f_{l,d,\text{imag}}(\theta_d) \approx f_{l,d-1,\text{imag}}(\theta^*_{d-1})$, bringing the total number of circuits to two. For small values of $\tau$, the error of this approximation is proportional to the modification $\delta \theta_{d-1}$ of the previous parameter. A smaller value of $\tau$ produces a smaller value of $\delta \theta_{d-1}$, hence the error of this approximation can be systematically reduced by decreasing $\tau$.

\section{Time-dependent variational principle and matrix inversion}
\label{a:TDVP}

The time-dependent variational principle (\textsc{tdvp}) can be derived in the following way.
Our goal is to time-evolve an Ansatz $|\psi(\boldsymbol{\theta})\rangle$ via the Schr\"{o}dinger equation:
\eq{SE}{
i \dv{}{t} \ket{\psi(\boldsymbol{\theta})} = \hat{H} \ket{\psi(\boldsymbol{\theta})} ,
}
where we have set $\hbar = 1$.
The right-hand side of equation~\eqref{eq:SE} may leave the variational space of $\ket{\psi(\boldsymbol{\theta})}$, which is created by all possible choices for $\boldsymbol{\theta}$. To stay in the variational space, we minimize the squared Euclidean distance:
\eq{TDVP}{
\text{dist}_{\text{TDVP}} &= \norm{i \dv{}{t} \ket{\psi(\boldsymbol{\theta})} - \hat{H}\ket{\psi(\boldsymbol{\theta})}}^2\\
&= \left( \dv{}{t}\bra{\psi(\boldsymbol{\theta})} \right) \left( \dv{}{t}\ket{\psi(\boldsymbol{\theta})} \right) + i \left( \dv{}{t}\bra{ \psi(\boldsymbol{\theta})} \right) \hat{H} \ket{\psi(\boldsymbol{\theta})} - i \bra{\psi(\boldsymbol{\theta})} \hat{H} \left( \dv{}{t}\ket{\psi(\boldsymbol{\theta})} \right) + \mel{\psi(\boldsymbol{\theta})}{\hat{H}^{2}}{\psi(\boldsymbol{\theta})} .
}
Using the chain rule:
\eq{TDVP_CR}{
\dv{}{t} \ket{\psi(\boldsymbol{\theta})} = \sum_{k} \frac{\partial \ket{\psi(\boldsymbol{\theta})}}{\partial \theta_{k}} \dv{\theta_{k}}{t} ,
}
and the definitions:
\eq{TDVP_DEF}{
B_{k} & := \dv{\theta_{k}}{t}\\
A_{j, k} & := \left( \frac{\partial \bra{\psi(\boldsymbol{\theta})}}{\partial \theta_{j}} \right) \left( \frac{\partial \ket{\psi(\boldsymbol{\theta})}}{\partial \theta_{k} } \right)\\
C_{j} & := \left( \frac{\partial \bra{ \psi(\boldsymbol{\theta})}}{\partial \theta_{j}} \right) \hat{H} \ket{\psi{\boldsymbol(\theta})} ,
}
we rewrite Eq.~\eqref{eq:TDVP} as:
\eq{TDVPNew}{
\text{dist}_{\text{TDVP}} &= \sum_{j, k} \bar{B}_{j} A_{j, k} B_{k} + i \sum_{j} \bar{B}_{j} C_{j} - i \sum_{j} \bar{C}_{j} B_{j} + \text{const.}
}
In a PQC Ansatz, the variational parameters are rotation angles. Thus, we now restrict $\theta_{j}$ to be real and obtain $B_{j}$ that are also real. That is, $\bar{B}_{j}= B_{j}$ in the equation above. The minimum of this equation in terms of the $B_{j}$ can be determined by taking the derivatives and equating them to zero:
\eq{TDVP_DER}{ 
\frac{\partial \text{dist}_{\text{TDVP}}}{\partial B_{j}} = \sum_{k} 2 \Re(A_{j, k}) B_{k} - 2 \Im(C_{j}) = 0 .
}
This is equivalent to:
\eq{TDVP_TD}{
\sum_{k} \Re(A_{j, k}) B_{k} = \Im(C_{j}) .
}
Notice that Eq.~\eqref{eq:TDVP_TD} can be written as a matrix vector equation by defining $\Re(A_{j, k})$ to be the matrix elements inside a matrix $A$, and defining $B_{k}$ to be the vector elements inside a vector $\boldsymbol{B}$, and defining $\Im(C_{j})$ to be the vector elements inside a vector $\boldsymbol{C}$. This leads to the final \textsc{tdvp} equation:
\eq{TDVPFinal}{
A \boldsymbol{B} = \boldsymbol{C} .
}
This is an equation for the time-dependence of the parameters in our PQC Ansatz since $\boldsymbol{B}=\dv{}{t} \boldsymbol{\theta}$. Therefore, \textsc{tdvp} replaces the original Schr\"{o}dinger equation~\eqref{eq:SE} with a new equation~\eqref{eq:TDVPFinal} that time-evolves the variational parameters directly. Although we focused here on real time evolution, other applications such as imaginary time evolution lead to similar systems of linear equations~\cite{YuEtAl19}.

To quantify the accuracy of our solution vector $\boldsymbol{B}$ in Eq.~\eqref{eq:TDVPFinal} it is important to emphasize that the matrix $A$ and the vector $\boldsymbol{C}$ are constructed from a finite number of measurements on a quantum computer and therefore have a finite precision~\cite{LiBe17, YuEtAl19}.
The \textsc{tdvp} algorithm determines the individual elements in $A$ and $\boldsymbol{C}$ as mean values over $m$ measurements of specific quantum circuits and this mean value computation has the error $\epsilon_{\text{MC}}$ of classical Monte Carlo sampling scaling like $O(1/\sqrt{m})$.
Therefore, using a number of measurements $m$ for each element in the matrix $A$ and vector $\boldsymbol{C}$, both are accurate only up to an error scaling like $O(1/\sqrt{m})$.
To analyze the effect of this error in $A$ on the solution of Eq.~\eqref{eq:TDVPFinal}, we replace $A$ by $A + \delta A$ where now $A$ represents the exact $A$ and $\delta A$ its error.
Similarly we replace $\boldsymbol{B}$ by $\boldsymbol{B} + \delta \boldsymbol{B}$:
\eq{ErrorA1}{
(A + \delta A) (\boldsymbol{B} + \delta \boldsymbol{B}) & = \boldsymbol{C}\\
A \delta \boldsymbol{B} + \delta A \boldsymbol{B} & = 0\\
\delta \boldsymbol{B} & = -A^{-1} \delta A \boldsymbol{B}\\
\norm{ \delta \boldsymbol{B} } & \leq \norm{ A^{-1} } \norm{\delta A} \norm{\boldsymbol{B}}\\
\frac{\norm{\delta \boldsymbol{B}}} {\norm{\boldsymbol{B}}} & \leq \norm{A} \norm{A^{-1}} \frac{\norm{\delta A}}{\norm{A}} ,
}
where we have used $A \boldsymbol{B} = \boldsymbol{C}$, ignored $\delta A \delta \boldsymbol{B}$, and where $\norm{\cdot}$ is a norm.
We observe that $\norm{A} \norm{A^{-1}}$ is the condition number. Thus $\kappa = \norm{A} \norm{A^{-1}} = \sigma_{\text{max}} / \sigma_{\text{min}}$ where $\sigma_{\text{max}}$ denotes the largest and $\sigma_{\text{min}}$ the smallest singular value of $A$.
We also observe that $\norm{\delta A}/\norm{A}$ is the relative error of $A$ which scales like $O(1/\sqrt{m})$.
Therefore, we have obtained an upper bound for the relative accuracy of $\boldsymbol{B}$:
\eq{ErrorA2}{
\epsilon_{B}^{\text{max}} & \propto \frac{\kappa}{\sqrt{m}} .
}
Equivalently, Eq.~\eqref{eq:ErrorA2} states that for a desired accuracy $\epsilon$ in $\boldsymbol{B}$ we need the number of measurements $m$ to scale like $O(\kappa^{2}/\epsilon^{2})$.
The same scaling is obtained for finite precision $\boldsymbol{C}$ by repeating the above calculation using $\boldsymbol{C} + \delta \boldsymbol{C}$.
We conclude that the matrix inversion required in the original \textsc{tdvp} methods~\cite{LiBe17, YuEtAl19} leads to a computational cost scaling like $O(\kappa^{2}/\epsilon^{2})$ for accuracy $\epsilon$.
Compared with the standard measurement error $O(1/\epsilon^{2})$ the additional factor of $\kappa^{2}$ can be computationally challenging especially for ill-conditioned matrices $A$.

\section{Numerical simulations}
\label{a:numerical}

For the numerical simulations we use the PQC Ansatz shown in Fig.~\ref{fig:PQCAndCones}~(a), with depth $\mathcal{D}=2$ and number of qubits $n \in \{4,6,8\}$ depending on the experiment. For expectations and Hadamard tests, we always use causal cones to reduce the size of the simulation. For one-qubit and two-qubit nearest-neighbor Trotter terms, the causal cones involve at most six qubits. This remains true if we include periodic boundary conditions in the Ansatz, such as the $U[\tfrac{n}{2},2]$ block shown in Fig.~\ref{fig:PQCAndCones}~(a). 

\begin{figure}
\centering
\includegraphics[width=124.605mm]{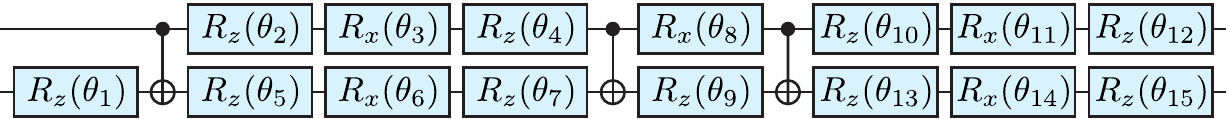}
\caption{The two-qubit block used in the simulations. Each block has three CNOTs and $N_p=15$ adjustable parameters consisting of angles of rotation about the canonical $x$ and $z$ axes.}
\label{fig:block_ansatz}
\end{figure}

Each block in the Ansatz is implemented using a two-qubit minimal construction proposed in Ref.~\cite{shende2004minimal} and shown in Fig.~\ref{fig:block_ansatz}. This construction has $N_p=15$ parameters and is universal for two-qubit unitaries up to a global phase. Note however that in all our experiments the phase is relevant since $n>2$.

For real time evolution, we use the following first- and second-order Trotter product formulas:
\begin{eqnarray}
 e^{-i t \hat{H}} & \approx & (e^{-i \tau \hat{H}_{X}} e^{-i \tau \hat{H}_{Z Z}})^{\frac{t}{\tau}}\\
 e^{-i t \hat{H}} & \approx & (e^{-i \tau \hat{H}_{Z Z} / 2} e^{-i \tau \hat{H}_{X}} e^{-i \tau \hat{H}_{Z Z} / 2})^{\frac{t}{\tau}},
\end{eqnarray}
respectively.
For small values of the time step $\tau$, the leading terms for the error per times step in the first- and second-order formulas are $\frac{\tau^{2}}{2} [\hat{H}_{X}, \hat{H}_{Z Z}]$ and $\frac{\tau^{3}}{12} ( \frac{1}{2} [\hat{H}_{Z Z}, [\hat{H}_{X}, \hat{H}_{ZZ}]] + \frac{1}{2} [[\hat{H}_{Z Z}, \hat{H}_{X}], \hat{H}_{ZZ}] + [\hat{H}_{X}, [\hat{H}_{X}, \hat{H}_{ZZ}]] +  [[\hat{H}_{Z Z}, \hat{H}_{X}], \hat{H}_{X}] )$, respectively.
Here we denote by $\hat{H}_{Z Z} = -J \sum_{j} \hat{\sigma}_{j}^{z} \hat{\sigma}_{j+1}^{z}$ and $\hat{H}_{X} = -J \lambda \sum_{j} \hat{\sigma}_{j}^{x}$ a decomposition of the quantum Ising Hamiltonian in Eq.~\eqref{eq:hamiltonian} into two noncommuting parts.
All terms inside $\hat{H}_{Z Z}$ and inside $\hat{H}_{X}$ commute with each other and therefore we use the additional decompositions:
\begin{eqnarray}
 e^{-i \tau \hat{H}_{Z Z}} & = & \prod_{j} e^{-i \tau J \hat{\sigma}_{j}^{z} \hat{\sigma}_{j+1}^{z}}\\
 e^{-i \tau \hat{H}_{X}} & = & \prod_{j} e^{-i \tau J \lambda \hat{\sigma}_{j}^{x}}.
\end{eqnarray}

Expectations and Hadamard tests are calculated exactly. We do not include finite-sampling noise or hardware noise. All numerical simulations are programed in \textsc{python} using the \textsc{qutip} library~\cite{johansson2012qutip}.

\twocolumngrid

\end{document}